\documentclass[11pt]{article}
\usepackage{latexsym, amsmath, amscd, amssymb, amsthm}
\usepackage[T1]{fontenc}
\usepackage{graphicx} 
\usepackage{pgf}

 \usepackage{xypic}

\def\N{{\mathbb N}}     
\def\R{{\mathbb R}}     
\def\Z{{\mathbb Z}}     

\newcommand{\Sp}{\mbox{\small $\mathcal S$}}
\newcommand{\Cx}{\mbox{\small $\mathcal C$}}
\newcommand{\Rmas}{\mbox{$\R_{> 0}$}}

\newcommand{\Znn}{\mbox{$\Z_{\geq 0}$}}
\newcommand{\Nnn}{\mbox{$\N_{\geq 0}$}}
\newcommand{\lap}{\mbox{$\mathcal L$}}
\newcommand{\og}{\overline{G}}

\newcommand{\rk}{\mbox{rk$\,$}}
\newcommand{\Null}{\mbox{$\mathcal N$}}

\newcommand{\ra}{\rightarrow}

\newtheorem{theorem}{Theorem}
\newtheorem{prop}{Proposition}
\newtheorem{cor}{Corollary}
\theoremstyle{definition}

\begin{document}

\title{Complex-linear invariants of biochemical networks}

\author
{R. L. Karp$^{1}$, M. P\'erez Mill\'an$^{2}$, T. Dasgupta$^{1}$, \\
A. Dickenstein$^{2,3}$, J. Gunawardena,$^{1}$\\
\\
\normalsize{$^1$Department of Systems Biology, Harvard Medical School}\\
\normalsize{Boston, MA, USA}\\
{\normalsize{{\tt Robert\_Karp, Tathagata\_Dasgupta, jeremy@hms.harvard.edu}}}\\
\normalsize{$^2$Departamento de Matem\'{a}tica, Universidad de Buenos Aires}\\
\normalsize{Buenos Aires, Argentina}\\
\normalsize{\tt mpmillan, alidick@dm.uba.ar}\\
\normalsize{$^3$IMAS-CONICET, C1428EGA Buenos Aires, Argentina}
}

\maketitle

\begin{abstract}
The nonlinearities found in molecular networks usually prevent mathematical analysis of network behaviour, 
which has largely been studied by numerical simulation. This can lead to difficult problems of parameter
 determination. However, molecular networks give rise, through mass-action kinetics, to polynomial dynamical 
systems, whose steady states are zeros of a set of polynomial equations. These equations may be analysed by 
algebraic methods, in which parameters are treated as symbolic expressions whose numerical values do not 
have to be known in advance.  For instance, an ``invariant'' of a network is a polynomial expression 
on selected state variables that vanishes in any steady state. Invariants have been found that encode 
key network properties and that discriminate between different network structures. 
Although invariants may be calculated by computational algebraic methods, such as Gr\"obner bases, 
these become computationally infeasible for biologically realistic networks. Here, we exploit 
Chemical Reaction Network Theory (CRNT) to develop an efficient procedure for calculating invariants 
that are linear combinations of ``complexes'', or the monomials coming from mass action. We show how 
this procedure can be used in proving earlier results of Horn and Jackson and of Shinar and Feinberg 
for networks of deficiency at most one. We then apply our method to enzyme bifunctionality, including 
the bacterial EnvZ/OmpR osmolarity regulator and the mammalian 6-phosphofructo-2-kinase/fructose-2,6-bisphosphatase
 glycolytic regulator, whose networks have deficiencies up to four. We show that bifunctionality leads to 
 different forms of concentration control that are robust to changes in initial conditions or total amounts. 
 Finally, we outline a systematic procedure for using complex-linear invariants to analyse molecular networks 
 of any deficiency.
\end{abstract}

\section{INTRODUCTION}

A molecular interaction network within a cell may be decomposed into elementary biochemical reactions, such as
\begin{equation}
2S_1 + 3S_2 \rightarrow 4S_3 \,,
\label{e-reac}
\end{equation}
where the $S_i$ are distinct chemical species. (This stoichiometry is unlikely but helpful for illustrative purposes.) Under mass-action kinetics, the rate of such a reaction is proportional to the concentrations of the substrates, taking stoichiometry into account. Hence,
\begin{equation}
\frac{dx_3}{dt} = 4\kappa_1x_1^2 x_2^3 \,,
\label{e-mon}
\end{equation}
where $x_i$ is the concentration of species $S_i$, $x_i = [S_i]$, and $\kappa_1 \in \Rmas$ is the positive mass-action rate constant. In a biochemical network, reactions contribute production and consumption terms consisting of monomials like $4\kappa_1x_1^2 x_2^3$ to the rates of formation of the species in the network. This results in a system of ordinary differential equations (ODEs), $dx/dt = f(x;\kappa)$, in which each component rate function $f_i(x;\kappa)$ is a polynomial in the state variables $x_1, x_2, \cdots, x_n \in \R$ and $\kappa_1, \cdots, \kappa_p \in \Rmas$ are positive rate constants.

The nonlinearities in {\bf(\ref{e-mon})} usually preclude mathematical analysis of the dynamical behaviour of 
such ODE systems, which are customarily studied by numerical simulation. This requires that the rate constants 
be given numerical values, which in most cases are neither known nor readily measurable. The resulting
 ``parameter problem'' remains a major difficulty in exploiting mathematical models, \cite{gun-misb-1}. 
 However, the steady states of such ODEs are zeros of a set of polynomial equations, $f_1(x,\kappa) = 0, 
 \cdots, f_n(x,\kappa) = 0$. Computational algebra and algebraic geometry provide powerful tools for 
 studying these solutions, \cite{clos}, 
and these have recently been used to gain new biological insights, \cite{cdss08,dcg10,rg08,psc11,mg09,mg07}. 
The rate constants can now be treated as symbolic parameters, whose numerical values do not need to be known 
in advance. 
The capability to rise above the parameter problem allows more general results to be obtained than can 
be expected from numerical simulation, \cite{mg07}. 

The focus on steady states, rather than transient dynamics, is still of substantial interest. For instance, in time-scale separation, which has been a widespread method of simplification in biochemistry and molecular biology, a fast sub-system is assumed to be at steady state with respect to a slower environment and steady-state analysis is used to eliminate the internal complexity in the sub-system, \cite{gun-mt}. Approximate or quasi-steady states have also been shown to exist under various cellular conditions and can now be engineered {\em in vivo}, \cite{lk99,kvbawf}.  Finally, steady states provide the skeleton around which the transient dynamics unfolds, so knowledge of the former can be helpful for understanding the latter.

The present paper focusses on the algebraic concept of an ``invariant'': a polynomial expression on selected state variables that is zero in any steady state, with the coefficients of the expression being rational expressions in the symbolic rate constants, \cite{rg08}. Recall that a rational expression is a quotient of two polynomials; an example of such being the classical Michaelis-Menten constant of an enzyme, \cite{cb95}. (A more general definition of an invariant allows the coefficients to include conserved quantities, \cite{xg11b}, but this extension is not discussed here.) Since each of the rate functions, $f_i(x;\kappa)$, is zero in any steady state, the force of the definition comes from the restriction to ``selected state variables''. It is possible that, by performing appropriate algebraic operations on $f_1, \cdots, f_n$, non-selected variables can be eliminated, leaving a polynomial expression on only the selected variables that must be zero in any steady state.

Invariants turn out to be surprisingly useful. They have been shown to characterise the biochemical networks underlying multisite protein phosphorylation, \cite{rg08}, suggesting that different network architectures can be identified through experimental measurements at steady state. If an invariant has only a single selected variable that appears linearly, this variable has the same value in any steady state since it is determined solely by the rate constants. In particular, its value is unaffected by changes to the initial conditions or to the total amounts of any species. This is ``absolute concentration robustness'' (ACR), as introduced in \cite{sf09}, which accounts for experimental findings in some bacterial bifunctional enzymes, \cite{bg03,smma07,sra09}. The mammalian bifunctional enzyme, 6-phosphofructo-2-kinase/fructose-2,6-bisphosphatase (PFK-2/FBPase-2), which has a more complex enzymatic network, also yields invariants, with implications for regulation of glycolysis, \cite{dcg10}. The methods developed here provide a systematic way to analyse such bifunctional enzymes, as explained below.

Computational algebra exploits the method of Gr\"obner bases to provide an Elimination Theorem, \cite{clos}, that permits variables to be systematically eliminated among the rate equations, $f_1, \cdots, f_n$, \cite{rg08}. Algorithms for calculating Gr\"obner bases are available in general-purpose tools like Mathematica, Matlab and Maple and in specialised mathematical packages such as Singular and Macaulay2\footnote{Available from {\tt www.singular.uni-kl.de} and {\tt www.math.uiuc.edu/Macaulay2}}. However, these algorithms are computationally expensive for the task at hand. They have been developed for general sets of polynomials and have not been optimised for those coming from biochemical networks. For instance, Mathematica's Gr\"obner basis algorithm does not terminate on the network for PFK-2/FBPase-2. If invariants are to be exploited further, alternative approaches are needed.

The nonlinearity in mass action comes from the pattern of substrate stoichiometry in {\bf(\ref{e-reac})}, which gives rise to the monomial nonlinearity in {\bf(\ref{e-mon})}. In the language of Chemical Reaction Network Theory (CRNT), the patterns of stoichiometry that appear on either side of a reaction arrow are called ``complexes'', \cite{fein79,gun03}. Reaction {\bf(\ref{e-reac})} has three species, $S_1, S_2$ and $S_3$ and two complexes, $2S_1 + 3S_2$ and $4S_3$. If $C = e_1S_1 + \cdots + e_nS_n$, where $e_i \in \Znn$ are nonnegative integer stoichiometries, then the complex $C$ gives the monomial, $x^C$, where $x^C = x_1^{e_1} \cdots x_n^{e_n}$. 

Aside from this nonlinearity, the defining rate equations come from linear processes on complexes. This observation is the starting point of CRNT and reveals that biochemical networks conceal much linearity behind their nonlinearity, \cite[and see below]{fein79,gun03}. This suggests the possibility of using fast linear methods, in preference to slow polynomial algorithms, to construct a subset of invariants: those that are symbolic linear combinations of the complex monomials, $x^C$. As before, this definition acquires substance by restricting the complexes that can appear. If $C_1, \cdots, C_k$ are the selected complexes, then a complex-linear invariant is a polynomial expression of the form $a_1x^{C_1} + \cdots + a_kx^{C_k}$, that is zero in any steady state, where $a_1, \cdots, a_k$ may be rational expressions in the symbolic rate constants. 

In this paper, we examine a large class of complex-linear invariants that we call ``type 1''. We determine the dimension of the space of type 1 invariants (Proposition~\ref{t-dim}) and provide a linear algorithm for calculating them (Theorem~\ref{t-inv}). We point out how invariants can be used in proving previous results of Horn and Jackson, \cite{hj72}, and the Shinar-Feinberg Theorem for ACR, \cite{sf09}. We then apply the method to contrast two examples of enzymatic bifunctionality, the bacterial EnvZ/OmpR osmolarity regulator and the mammalian PFK-2/FBPase-2 glycolytic regulator. The method is sufficiently straightforward that the invariants for networks of this kind can be found by manual inspection of an appropriate matrix. Finally, we outline a systematic procedure for analysing any network using type 1 complex-linear invariants. An appendix contains supplementary information.

\section{RESULTS}

\subsection{Background on CRNT} \hspace{0.5em} It is assumed that there are $n$ species, 
$\Sp = \{S_1, \cdots, S_n\}$, whose concentrations are $x_1, \cdots, x_n \in \R$, respectively, 
and $m$ complexes, $\Cx = \{C_1, \cdots, C_m\}$. Complexes are regarded formally as multisets of species, 
$C_i \in \N^{\Sp}$, where the value of the multiset, $C_i$, on the species $S_j$, denoted $C_i(S_j)$ is the
 stoichiometry of $S_j$ in $C_i$. Accordingly, $C_i = C_i(S_1)S_1 + \cdots + C_i(S_n)S_n$. The reactions 
 in the network define a directed graph on the complexes, with an edge $C_i \ra C_j$ whenever there is a 
 reaction with substrate stoichiometry given by $C_i$ and product stoichiometry given by $C_j$. 
 The corresponding mass-action rate constant gives each edge a label, $\kappa_{i,j}$, treated as 
 a positive symbol, $\kappa_{i,j} \in \Rmas$. 

Any directed graph, $G$, with labels in $\Rmas$ gives rise to an abstract dynamics in which each edge is treated as if it were a first-order chemical reaction with its label as rate constant. Since the rates are all first-order, the dynamics are linear and may therefore be written in matrix terms as $dy/dt = \lap(G).y$, where $y \in \R^m$ is a column vector, consisting of an abstract concentration $y_i$ at each node $i$ of $G$, and $\lap(G)$ is a $m \times m$ matrix called the Laplacian matrix of $G$. Here, ``.'' signifies matrix multiplication, regarding vectors as matrices of one row or one column. The graph Laplacian has wide application in biology, as described in \cite{gun-mt}, where additional information may be found.

In CRNT, the Laplacian, $\lap(G): \R^m \ra \R^m$, provides a linear analogue for complexes of the nonlinear function, $f:\R^n \ra \R^n$, for species, in the following sense. Let $\Psi: \R^n \ra \R^m$ be the nonlinear function that lists the monomials for each complex, $\Psi(x) = (x^{C_1}, \cdots, x^{C_m})^\dagger$. Here, $^\dagger$ denotes transpose. Let $Y: \R^m \ra \R^n$ be the linear function that associates to each complex, considered as a basis element of $\R^m$, its corresponding stoichiometry pattern; the $i$-th column of the resulting matrix is then $(C_i(S_1), \cdots, C_i(S_n))^\dagger$. With these definitions, it may be checked that $f(x) = Y.\lap(G).\Psi(x)$, for any $x \in \R^n$, as depicted in the commutative diagram in the top left of Figure~\ref{f-0}.

This fundamental decomposition is due to Horn and Jackson, \cite{hj72}, and is the starting point of CRNT. They did not use the Laplacian description, which was introduced in \cite{cdss08} and exploited further in \cite{gun-mt,mg09}. To analyse steady states, where $f(x) = 0$, it is particularly useful to know the kernel of $\lap(G)$: $\ker\lap(G) = \{ y \in \R^m\;|\; \lap(G).y = 0 \}$. This was first determined by Feinberg and Horn, \cite[Appendix]{fh77}, by a non-constructive method. Here, we briefly describe the constructive method introduced in \cite{gun-mt}, which shows how $\ker\lap(G)$ can be algorithmically calculated from $G$.

This can be done in two stages, \cite{gun-mt}. First, if $G$ is ``strongly connected'', so that any two distinct nodes are linked by a contiguous series of edges in the same direction, then $\dim\ker\lap(G) = 1$. The Matrix-Tree Theorem provides an explicit construction of a basis element, $\rho_G \in \R^m$, in terms of the spanning trees of $G$: $\ker\lap(G) = \langle\,\rho_G\,\rangle$. The components $(\rho_G)_i$ are polynomials in the symbolic labels but the details of their calculation are not needed here. If $G$ is not strongly connected, it can be partitioned into its maximal strongly-connected sub-graphs, or ``strongly connected components'' (SCCs). These inherit from $G$ a directed graph structure, $\og$, in which there is an edge in $\og$ from SCC $G_u$ to SCC $G_v$ whenever there is an edge in $G$ from some node in $G_u$ to some node in $G_v$. $\og$ cannot have any directed cycles and so always has terminal SCCs, with no edges leaving them. Let these be $G_1, \cdots, G_T$. For each $1 \leq t \leq T$, let $\rho^t \in \R^m$ be the vector which, for vertices of $G$ that lie in $G_t$, agrees with the vector $\rho_{G_t}$, coming from the Matrix-Tree Theorem applied to $G_t$ as an isolated graph, and, for all other vertices, $j$, $(\rho^t)_j = 0$. Then, the $\rho^t$ form a basis for $\ker\lap(G)$:
\begin{equation}
\ker\lap(G) = \langle\, \rho^1, \cdots, \rho^T \,\rangle \,.
\label{e-klap}
\end{equation}
This gives algebraic expressions for the components of the basis vectors as polynomials in the symbolic labels. Note that $\rho^t$ may be very sparse, being non-zero only for vertices in the single SCC $G_t$.

\subsection{Generating complex-linear invariants}
\label{s-gen}

Depending on the application, invariants may be required that involve only certain complexes, $C_{i_1}, \cdots, C_{i_k}$, for instance, those involving species with more easily measurable concentrations. Since the indices can be permuted so that the complexes of interest appear first in the ordering, it can be assumed that invariants are sought on $C_1, \cdots, C_k$. Let $M$ be the $n \times m$ matrix representing the linear part of the CRNT decomposition, $M = Y.\lap(G)$. A simple way to construct a complex-linear invariant on $C_1, \cdots, C_k$ is to find a vector, $a^\dagger \in \R^k$, such that, if $(a,0)^\dagger \in \R^m$ is $a$ extended with $m-k$ zeros, $(a,0) = (a_1, \cdots, a_k, 0, \cdots, 0)$, then $(a,0)$ is in the rowspan of $M$. That is, it is a linear combination of the rows of $M$. If $x \in \R^n$ is any steady state of the system, so that $f(x) = 0$, then $\Psi(x) \in \ker M$ because $M.\Psi(x) = Y.\lap(G).\Psi(x) = f(x) = 0$. Since $(a,0)$ is in the rowspan of $M$, $(a,0).\Psi(x) = 0$. Hence, by definition of $\Psi$, $a_1x^{C_1} + \cdots + a_kx^{C_k} = 0$, giving a complex-linear invariant on $C_1, \cdots, C_k$.

Not all such invariants may arise in this way. For that to happen, it is necessary not just for $(a,0).\Psi(x) = 0$ whenever $x$ is a steady state but for $(a,0).v = 0$ for all $v \in \ker M$. The relationship between $\ker M$ and $\{\Psi(x) \;|\; \mbox{$x$ is a steady state}\}$ is not straightforward. To sidestep this problem, we focus here only on those invariants, $a_1x^{C_1} + \cdots + a_kx^{C_k}$, in which $(a,0)$ is in the rowspan of $M$. We call these type 1 complex-linear invariants. Non-type 1 invariants do exist, as we show in the Supporting Information (SI). The type 1 invariants form a vector space that we abbreviate $I_k$; note that $I_k$ depends on $C_1, \cdots, C_k$ and not just on $k$. Two basic problems are, first, to determine the dimension of $I_k$ and, second, to generate its elements. 

A simple solution to the second problem is to break the matrix $M$ into the $n \times k$ sub-matrix $K$ consisting of the first $k$ columns of $M$ and the $n \times (m-k)$ sub-matrix $N$ consisting of the remaining $m-k$ columns, so that $M = K\,|\,N$. Any vector $b^\dagger \in \R^n$ which is in the left null space of $N$, $b \in \Null_L(N)$, so that $b.N = 0$, gives an $(a,0) = b.M$ that is in the rowspan of $M$. The assignment $b \ra b.M$ thereby defines a surjection, $\Null_L(N) \ra I_k$. Moreover, $b_1.M = b_2.M$, if, and only if, $(b_1 - b_2) \in \Null_L(M) \subseteq \Null_L(N)$. Hence, there is an isomorphism $I_k \cong \Null_L(N)/\Null_L(M)$. If $X$ is any $n \times r$ matrix, $\dim\Null_L(X) = n - \rk X$. We conclude that $\dim I_k = \rk M - \rk N$, which yields an efficient way to determine the dimension of $I_k$ by Gaussian elimination.

In principle, this provides an automatic procedure for identifying subsets of complexes with non-trivial invariants. First determine $\rk M$ and then, for each subset $Z \subseteq \{1, \cdots, m\}$, determine the rank of the submatrix of $M$ formed by those columns not in $Z$. If the latter is smaller than the former, then there are non-trivial type 1 complex-linear invariants on the complexes in $Z$. In practice, it is usually more efficient to use biological knowledge of the example being studied and the question being asked to narrow the choice of $Z$. We outline such a systematic procedure for finding invariants in the last section.

The method above amounts to eliminating the complexes $C_{k+1}, \cdots, C_m$ by taking linear combinations of the defining rate functions, $f_1, \cdots, f_n$. This can be biologically informative because it suggests which rate functions, and, hence, which species at steady state, determine the invariant, \cite{dcg10}. 

\subsection{Duality and the structure of $I_k$}
\label{s-du}

\begin{figure}
\centering 
\includegraphics[viewport=54 238 574 726,width=0.8\textwidth,height=\textheight,keepaspectratio]{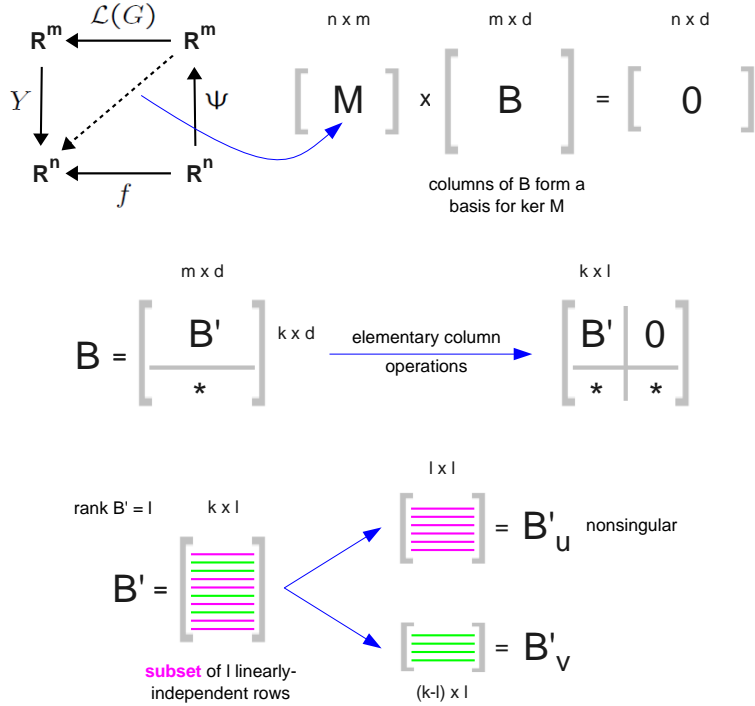}
\caption{Schematic illustration of the dual procedure for calculating type 1 complex-linear invariants using (\ref{e-inv}). The fundamental decomposition of CRNT is shown in the commutative diagram in the top left corner, for which $f(x) = Y.\lap(G).\Psi(x)$. The matrix $M$ corresponds to the linear part of this decomposition, $M = Y.\lap(G)$. The matrix $B$ has columns consisting of a basis for $\ker M$. The matrix $B'$ consists of the first $k$ rows of $B$. Assuming that $\rk B' = l$, elementary column operations can be applied to $B$ to bring it into lower-triangular block form. By a mild abuse of notation, the upper-left block continues to be called $B'$. Finally, a subset, $U$, of linearly independent rows (magenta) of $B'$ yields the nonsingular matrix $B'_U$, while the subset $V$ of remaining rows (green) yields the matrix $B'_V$, from which type 1 complex-linear invariants can be calculated using (\ref{e-inv}). \label{f-0}}
\end{figure}

In this section, we present an alternative procedure for calculating complex-linear invariants, which is based on duality and exploits the sparsity of {\bf (\ref{e-klap})}. The procedure is schematically illustrated in Figure~\ref{f-0}, as an aid to following the details. 

We start, as before, with the $n \times m$ matrix, $M$ that represents the linear part of the CRNT decomposition, $M = Y.\lap(G)$. Let $d = \dim\ker M$ and let $B$ be any $m \times d$ matrix whose columns form a basis of $\ker M$. Then, $M.B = 0$ and the rowspan of $M$ and the columnspan of $B$ are dual spaces of each other. If $a^\dagger \in \R^k$, then $(a,0)$ is in the rowspan of $M$ if, and only if, $(a,0).B = 0$. If $B'$ is the $k \times d$ sub-matrix of $B$ consisting of the first $k$ rows, then $(a,0).B = 0$ if, and only if, $a.B' = 0$. Hence, type 1 invariants form the dual space to the columns of $B'$. 

\begin{prop}\label{prop:1}
The space $I_k$ of type 1 complex-linear invariants on the complexes $C_1, \cdots, C_k$ satisfies $\dim I_k = \rk M - \rk N = k - \rk B'$. 
\label{t-dim}
\end{prop}

Let $l = \rk B'$. Note that $l \leq \min(k,d)$. If $l = k$, then $\dim I_k = 0$ and there are no type 1 invariants on $C_1, \cdots, C_k$. If, however, $l < k$, then the original matrix $B$ can be simplified in two steps. First the columns. Since the column rank of $B'$ is $l$, elementary column operations---interchange of two columns, multiplication of a column by a scalar, addition of one column to another---can be applied to the columns of $B'$, to bring the last $d-l$ columns to zero. If exactly the same elementary column operations are applied to the full matrix $B$, a new matrix is obtained, which we still call $B$, whose columns still form a basis for $\ker M$. $B$ is now in lower-triangular block form,
\begin{equation}
B =      \left(\begin{array}{c|c}
         B' & 0 \\
         \hline 
           *     & \vspace{0.2em}*
         \end{array}\right)
\label{e-bf}
\end{equation}
where, as before, $B'$ is the $k \times l$ sub-matrix consisting of the first $k$ rows and $l$ columns.

For the rows, since the row rank of $B'$ is still $l$, there are $l$ rows of $B'$ that are linearly independent. Let $U \subseteq \{1, \cdots, k\}$ be the corresponding subset of $l$ indices and let $V \subseteq \{1, \cdots, k\}$ be the subset of $k-l$ remaining indices. This defines a partition of the row indices of $B'$: $U \cap V = \emptyset$ and $U \cup V = \{1, \cdots, k\}$. Let $B'_U$ be the $l \times l$ sub-matrix of $B'$ consisting of the rows with indices in $U$ and $B'_V$ be the $(k-l) \times l$ sub-matrix consisting of the remaining rows of $B'$. Using the same notation for $a^\dagger \in \R^k$, $a.B' = 0$ if, and only if, $a_U.(B'_U) + a_V.(B'_V) = 0$. Since, by construction, $B'_U$ has full rank and is hence invertible, this may be rewritten as 
\begin{equation}
a_U = - a_V.(B'_V).(B'_U)^{-1} \,.
\label{e-au}
\end{equation}
This gives a non-redundant procedure for generating all elements of $I_k$ by choosing $a_V^\dagger \in \R^{k-l}$ arbitrarily and $a_U^\dagger \in \R^l$ to satisfy {\bf(\ref{e-au})}. The resulting $a^\dagger \in \R^k$ satisfy $a.B' = 0$ and give exactly the type 1 complex-linear invariants on $C_1, \cdots, C_k$. 

Using the same notation for $\Psi(x) \in \R^m$, the invariants themselves are given by $a_U.\Psi(x)_U + a_V.\Psi(x)_V = 0$, for any steady state $x \in \R^n$. Substituting {\bf(\ref{e-au})} and rearranging gives $a_V.(\Psi(x)_V - (B'_V).(B'_U)^{-1}.\Psi(x)_U) = 0$. Since $a_V$ can be chosen arbitrarily in the dual space, we conclude that
\begin{equation}
\Psi(x)_V = (B'_V).(B'_U)^{-1}.\Psi(x)_U \,,
\label{e-inv}
\end{equation}
which we summarise as follows.

\begin{theorem}\label{th:1}
Each of the $k-l$ rows of the matrix equation in {\bf(\ref{e-inv})} gives an independent type 1 complex-linear invariant on $C_1, \cdots, C_k$.
\label{t-inv}
\end{theorem}

This procedure relies on the choice of basis elements for $\ker M$ that make up the columns of $B$ and on the choice of the subset, $U$, of linearly independent rows of $B'$. These choices are not critical; different ones yield different bases for $I_k$. Up to linear combinations, the same invariants are found irrespective of the choices. All the calculations required are linear and can be readily undertaken in any computer algebra system with the rate constants treated as symbols. (Mathematica was used for the calculations in the SI.) The coefficients are then rational expressions in the symbolic rate constants.

\subsection{Haldane relationships and the Shinar-Feinberg Theorem}

Since $M = Y.\lap(G)$, $\ker M$ contains the subspace $\ker\lap(G)$. The structure of the latter is known from {\bf (\ref{e-klap})}. This should assist in the calculation of invariants, especially when $\ker\lap(G)$ is close to $\ker M$. We discuss two instances of this, which illustrate how complex-linear invariants are related to previous studies. 

Define the ``dynamic deficiency'' of a biochemical network, $\delta_D \in \Nnn$, to be the difference in dimension between the two subspaces: $\delta_D = \dim\ker M - \dim\ker\lap(G)$, or, equivalently, $\delta_D = \dim(\ker Y \cap \mbox{Image}\ \lap(G))$. This is different from the ``deficiency'' as usually defined in CRNT, \cite{fein79,gun03}, which we call the ``structural deficiency'', $\delta_S \in \Nnn$. While $\delta_D$ may depend on the values of rate constants, $\delta_S$ is independent of them. However, the former is more convenient for our purposes. 

It is known that $\delta_D \leq \delta_S$. Furthermore, if there is only a single terminal SCC in each connected component of $G$, which holds for the graph in Figure~\ref{f-1}C but not for that in Figure~\ref{f-3}A, then $\delta_D = \delta_S$, \cite{fein79,gun03}. Recall that a graph is connected if any two distinct nodes are linked by a path of contiguous edges, ignoring directions. A connected component of $G$ is then a maximal connected sub-graph. Distinct connected components are totally disconnected, with no edges between them. 

Suppose first that $\delta_D = 0$ and that there is a positive steady state $x \in (\Rmas)^n$. Since $\lap(G).\Psi(x) = 0$, $x$ is a ``complex-balanced'' steady state, in the terminology of Horn and Jackson, \cite{hj72}. According to {\bf(\ref{e-klap})}, the vectors $\rho^t$ provide a basis for $\ker M = \ker\lap(G)$ and, furthermore, $(\rho^t)_j \not= 0$ if, and only if, $C_j \in G_t$. Choose any terminal SCC of $G$, which we may suppose to be $G_1$, and suppose that $C_1, \cdots, C_k$ are the complexes in $G_1$. Choose the matrix $B$ so that $\rho^1$ is its first column and the other $\rho^t$ for $t > 1$ are assigned to columns arbitrarily. By construction, $B$ is already in lower-triangular block form and $l = 1$. Setting $U = \{1\}$ and $V = \{2, \cdots, k\}$ the $k-1$ type 1 invariants coming from {\bf(\ref{e-inv})} are $x^{C_i} = ((\rho^1)_i/(\rho^1)_1)x^{C_1}$ for $2 \leq i \leq k$. It is not difficult to see from the structure of $B$ that these are the only type 1 invariants. 

These invariants may be rewritten $x^{C_i}/x^{C_1} = (\rho^1)_i/(\rho^1)_1$ to resemble the Haldane relationships that hold between substrates and products of a reaction at equilibrium, \cite{cb95}. It follows from the construction of $\rho^t$ by the Matrix-Tree Theorem that the right-hand side of this relationship is determined by the rate constants, as expected for a Haldane relationship, \cite{cb95}. Horn and Jackson introduced the concept of a complex-balanced steady state, in part, to recover such generalised Haldane relationships for networks of reactions that might be in steady state but not at thermodynamic equilibrium, \cite{hj72,gun-mt}. 

Now suppose that $\delta_D = 1$. Then, $\ker M = \langle \,\chi, \rho^1, \cdots, \rho^T\,\rangle$, where $\chi \in \R^m$ is any vector in $\ker M$ that is not in $\ker\lap(G)$. Choose $B$ to have columns in the same order. Suppose that there are $k$ complexes that are not in any terminal SCC and that indices are chosen so that these are $C_1, \cdots, C_k$. Then, $(\rho^t)_i = 0$ for $1 \leq i \leq k$ and $1 \leq t \leq T$, so that $B$ is already in lower-triangular block form with $l = 1$. If $x \in (\Rmas)^n$ is a positive steady state, then $\Psi(x) \in \ker M$ and $\Psi(x)_i \not= 0$ for $1 \leq i \leq m$. If follows that $\chi_i \not= 0$ for $1 \leq i \leq k$. We may therefore choose $U = \{1\}$ and $V = \{2, \cdots, k\}$ and deduce from {\bf(\ref{e-inv})} that $x^{C_i} = (\chi_i/\chi_1)x^{C_1}$ for $2 \leq i \leq k$. 

These type 1 complex-linear invariants lead to the Theorem of Shinar and Feinberg on ACR, \cite{sf09}. Suppose that the structural deficiency of a network satisfies $\delta_S = 1$. Suppose further that $C_1$ and $C_2$ are two complexes that are not in any terminal SCC, whose stoichiometry differs only in species $S_q$. Since $\delta_D \leq \delta_S$ it must be that either $\delta_D = 0$ or $\delta_D = 1$. Suppose the former. The $\rho^t$ then form a basis for $\ker M$. Because $C_1$ is not in any non-terminal SCC, $v_1 = 0$ for any $v \in \ker M$. However, $\Psi(x) \in \ker M$ and, since $x \in (\Rmas)^n$, $\Psi(x)_1 \not= 0$. This contradiction shows that $\delta_D = 1$. It then follows from the invariant above that $(x_q)^{C_2(S_q)-C_1(S_q)} = \chi_2/\chi_1$. Hence, the steady-state concentration of $S_q$ depends only on the rate constants and not on the initial conditions or the total amounts and thereby exhibits ACR, \cite{sf09}.

\subsection{Bifunctional enzymes}\label{ss:be}

The previous calculations only exploited Theorem~\ref{t-inv} when $l = 1$. We now consider examples with $l > 1$. Details of the calculations are given in the SI. The examples concern enzyme bifunctionality. Enzymes are known for being highly specific but some exhibit multiple activities. One form of this arises when a protein catalyses both a forward phosphorylation---covalent addition of phosphate, with ATP as the donor---and its reverse dephosphorylation---hydrolysis of the phosphate group. What advantage does such bifunctionality bring over having two separate enzymes?

We discuss one bacterial and one mammalian example. In {\em Escherichia coli}, osmolarity regulation is implemented in part by the EnvZ/OmpR two-component system (Figure~\ref{f-1}A); for references, see \cite{smma07}. Here, the sensor kinase, EnvZ, autophosphorylates on a histidine residue and catalyses the transfer of the phosphate group to the aspartate residue of the response regulator, OmpR, which then acts as an effector. Bifunctionality arises because EnvZ, when ATP is bound, also catalyses hydrolysis of phosphorylated OmpR-P.

It was suggested early on that the unusual design of the EnvZ/OmpR system might keep the absolute concentration of OmpR-P stable, \cite{rs93}. This was later supported by experimental and theoretical analysis, \cite{bg03}, and the theoretical analysis was extended to other bifunctional two-component systems, \cite{smma07}. These {\em ad-hoc} calculations were clarified when a core network for EnvZ/OmpR was found to have $\delta_S = 1$ and the Shinar-Feinberg Theorem could be applied to confirm ACR for OmpR-P, \cite{sf09}. Attempts were made to broaden the analysis by extending the core network to include additional reactions thought to be present. For instance, EnvZ bound to ADP may also dephosphorylate OmpR-P. Adding these reactions to the core gives a network (Figure~\ref{f-1}B) with $\delta_S = 2$, so that Shinar-Feinberg can no longer be applied. However, it was shown by direct calculation in \cite[Supplementary Information]{smma07} that this network also satisfies ACR for OmpR-P.

Here, we use complex-linear invariants to confirm ACR and to find a formula for the absolute concentration value of OmpR-P in terms of the rate constants. The labelled, directed graph on the complexes has thirteen nodes and fifteen edges (Figure~\ref{f-1}C). Each connected component has only a single terminal SCC and $\delta_D = \delta_S = 2$. We can apply Theorem~\ref{t-inv} to systematically find two new invariants. 

\begin{cor}\label{cor:1}
If the complexes in the reaction network in Figure~\ref{f-1}B are ordered as shown in Figure~\ref{f-1}D, then the space of type 1 complex-linear invariants on the complexes $C_1, C_3, C_8, C_{11}$ has dimension 2 and the following are independent invariants,
\[\begin{array}{c}
\left(\frac{\textstyle k_1k_3}{\textstyle k_2}\right) x^{C_1} - (k_4 + k_5)x^{C_3} = 0 \\[0.6em]
k_5 x^{C_3} - \left(\frac{\textstyle k_{12}k_{10}}{\textstyle k_{11}+k_{12}}\right)x^{C_8} - \left(\frac{\textstyle k_{15}k_{13}}{\textstyle k_{14}+k_{15}}\right)x^{C_{11}} = 0\,.
\end{array}\]
\label{t-envi}
\end{cor}

Using the expressions for the complexes in Figure~\ref{f-1}D, it can be seen that 
\[ x^{C_8} = x^{C_3}[\mbox{OmpR-P}] \,,\hspace{1em} x^{C_{11}} = x^{C_1}[\mbox{OmpR-P}] \,.\]
Provided that $[\mbox{EnvZ-ATP}] = x^{C_3} \not= 0$, the invariants can be combined and simplified to yield the following expression
\begin{equation}
\begin{split}
& [\mbox{OmpR-P}] = \\ 
& \frac{k_1k_3k_5(k_{11}+k_{12})(k_{14}+k_{15})}{k_1k_3k_{10}k_{12}(k_{14}+k_{15}) + k_2k_{13}k_{15}(k_4+k_5)(k_{11}+k_{12})} \,.
\end{split}
\label{e-omp}
\end{equation}
This confirms that, as long as there is a positive steady state, the steady-state concentration of OmpR-P is not affected by changes in either the amount of OmpR or of EnvZ. The network exhibits ACR for OmpR-P, with the absolute value being given in terms of the rate constants by {\bf (\ref{e-omp})}.

We now turn to our second example. 6-Phosphofructo-1-kinase (PFK-1) is one of the key regulatory enzymes in glycolysis, converting the small molecule fructose-6-phosphate to fructose-1,6-bisphosphate (Figure~\ref{f-2}A); for references, see \cite{dcg10}. In mammalian cells, the bifunctional PFK-2/FBPase-2 has two domains. PFK-2 has the same substrate as PFK-1 but produces fructose-2,6-bisphosphate. This is a terminal metabolite that is not consumed by other metabolic processes. Instead, it acts as an allosteric effector, activating PFK-1 and inhibiting fructose-1,6-bisphosphatase, the reverse enzyme present in gluconeogenic cells, such as hepatocytes. The other domain, FBPase-2, catalyses the dephosphorylation of F2,6BP and produces F6P.

Biochemical studies lead to the reaction network in Figure~\ref{f-2}B. 
The kinase domain has an ordered, sequential mechanism and the kinase and 
phos\-pha\-tase domains operate simultaneously; for more details, see \cite{dcg10}. 
The corresponding labelled, directed graph on the complexes has fourteen nodes 
and nineteen edges (Figure~\ref{f-3}A). One of the connected components has two 
terminal SCCs, $\delta_D = 4$ and $\delta_S = 5$. 

\begin{cor}\label{cor:2}
If the complexes in the reaction network in Figure~\ref{f-2}B are ordered as shown in Figure~\ref{f-3}B, then the space of type 1 complex-linear invariants on the complexes $C_1, C_2, C_4, C_6, C_8, C_{11}$ has dimension 2 and the following are the independent invariants,
\[\begin{array}{c}
k_1x^{C_1} - k_2x^{C_2} + (k_{10}-k_{8})x^{C_6} - (k_9 + k_{11})x^{C_8} - k_{19}x^{C_{11}} = 0 \\[0.4em]
k_5 x^{C_4} - k_8x^{C_6} - k_{11}x^{C_8} + (k_{18}-k_{19})x^{C_{11}} = 0 \,.
\end{array}\]
\label{t-pfk2}
\end{cor}

The second invariant in Corollary~\ref{t-pfk2} was originally discovered by {\em ad-hoc} algebraic calculation. It is used in \cite{dcg10} to show that, if the kinase dominates the phosphatase, in the sense that $k_{18} > k_{19}$, then the steady state concentration of F6P is held below a level that depends only on the rate constants and not on the amounts of the enzymes or the substrate. Conversely, if the phosphatase dominates the kinase, so that $k_{18} < k_{19}$, then the steady state concentration of F2,6BP is similarly constrained below a level that depends only on the rate constants and not on the amounts. Interestingly, regulation of PFK-2/FBPase-2 by phosphorylation, under the influence of the insulin and glucagon, causes the kinase and phosphatase activities to be shifted between the regimes $k_{18} > k_{19}$ and $k_{18} < k_{19}$. The implications of this for control of glycolysis are discussed in \cite{dcg10}. 

\begin{figure}
\centering 
\includegraphics[viewport=20 10 754 550,width=0.8\textwidth,height=\textheight,keepaspectratio]{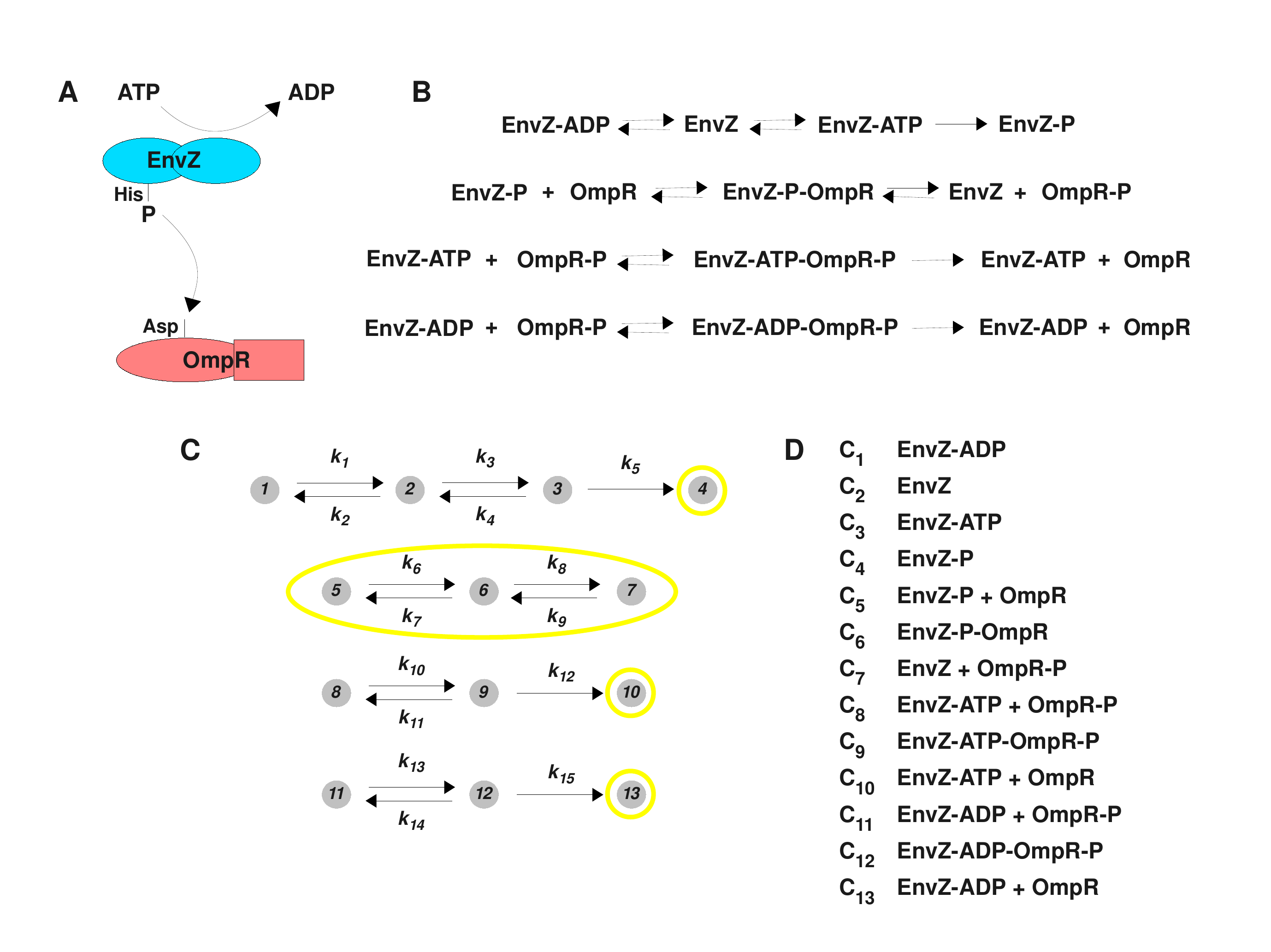}
\caption{Two component signalling and the {\em E. coli} osmolarity network. {\bf A} Schematic of two-component phospho-transfer between a histidine residue on the autophosphorylating sensor kinase (light blue) and an aspartate on the response regulator (red). {\bf B} Extended reaction network for the EnvZ/OmpR two-component osmoregulator in {\em E. coli}, following \cite[Supplementary Information]{smma07}. Hyphens, as in EnvZ-ATP, indicate the formation of a biochemical complex between the components. {\bf C} Corresponding labelled, directed graph on the complexes, with the terminal strongly connected components outlined in yellow. Each connected component has only a single terminal SCC. {\bf D} Numbering scheme for the complexes. \label{f-1}}
\end{figure}

\begin{figure}
\centering 
\includegraphics[viewport=30 10 618 556,width=0.8\textwidth,height=\textheight,keepaspectratio]{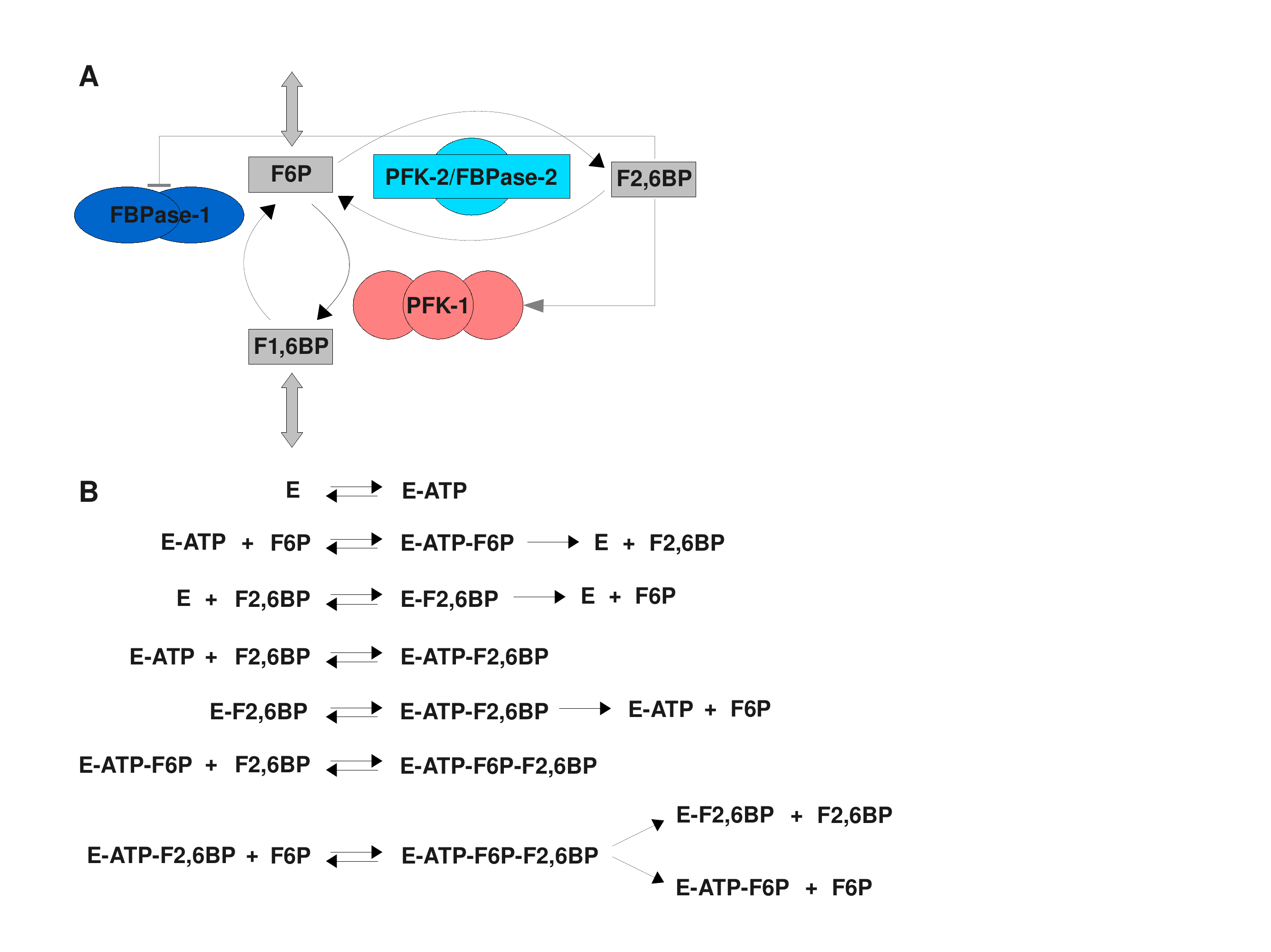}
\caption{The bifunctional enzyme 6-phosphofructo-2-kinase/fructose-2,6-bisphosphatase (PFK-2/FBPase-2). {\bf A} Schematic of the glycolysis/gluconeogenic pathway (broad gray arrows) at the step involving 6-phosphofructo-1-kinase (PFK-1, in red), that converts fructose-6-phosphate (F6P) into fructose-1,6-bisphosphate (F1,6BP), and fructose-1,6-bisphosphatase (FBPase-1, in dark blue), that catalyses the opposing reaction in gluconeogenic tissues. PFK-2/FBPase-2 (light blue) operates bifunctionally to produce and consume fructose-2,6-bisphosphate (F2,6BP), which allosterically regulates PFK-1 and FBPase-1, as shown. {\bf B} The corresponding reaction network. \label{f-2}}
\end{figure}

\begin{figure}
\centering 
\includegraphics[viewport=36 26 614 472,width=0.8\textwidth,height=\textheight,keepaspectratio]{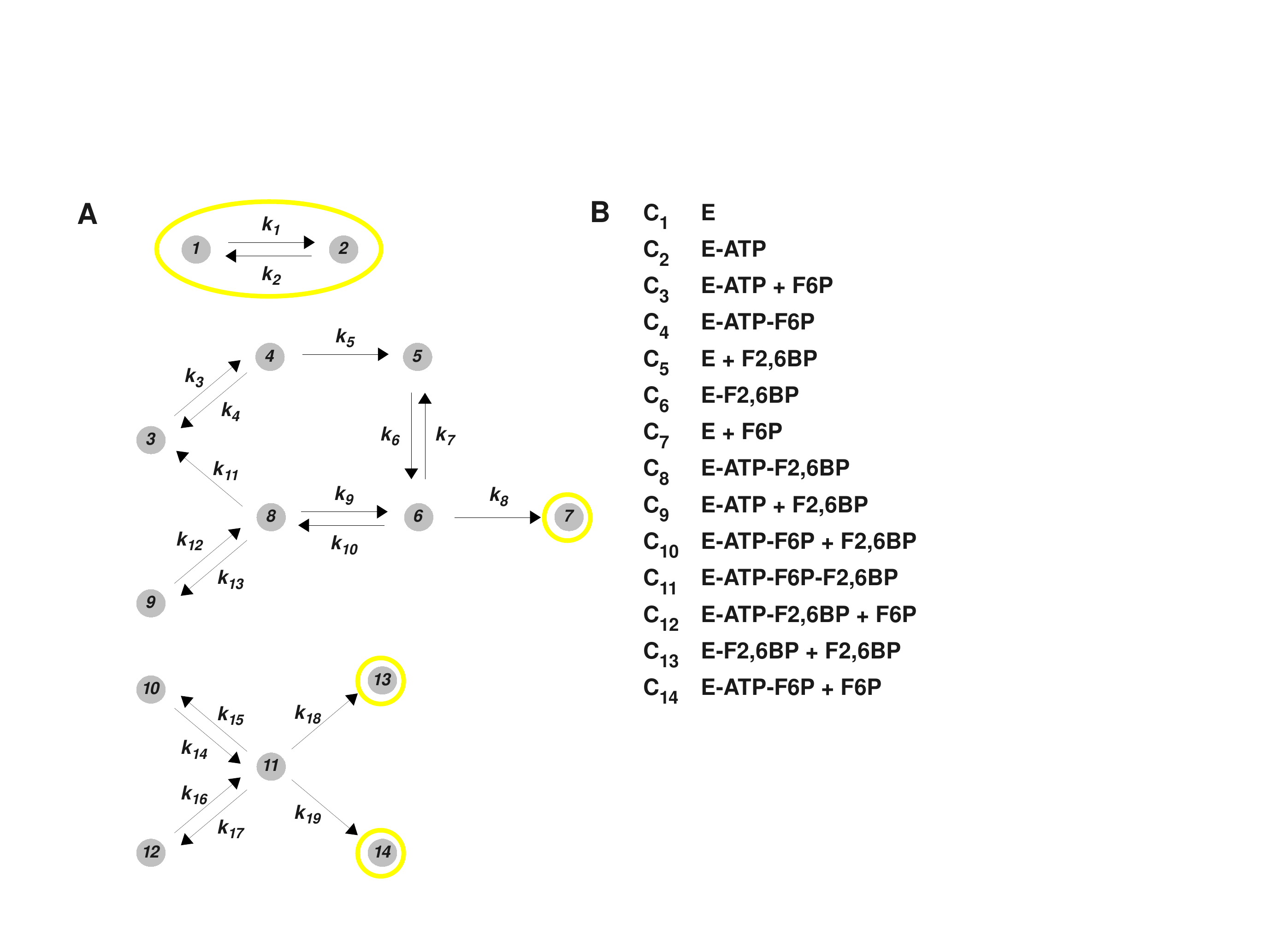}
\caption{PFK-2/FBPase-2, as in Figure~\ref{f-2}, in terms of complexes. {\bf A} Labelled, directed graph on the complexes, with the terminal SCCs outlined in yellow. The last connected component has two terminal SCCs. {\bf B} Numbering scheme for the complexes. \label{f-3}}
\end{figure}

\subsection{A systematic procedure}
\label{s-sys}

The two examples discussed above had already been analysed by other methods, so we had an idea of which invariants to expect and which subset of complexes to consider. For a new network, such information may not be available, so how can non-trivial type 1 complex-linear invariants (simply, ``invariants'') be found? The automatic procedure outlined in \S\ref{s-gen} can be used in principle but this becomes computationally infeasible when there are many complexes. We have found the following systematic procedure to be helpful on several examples. 

First determine the matrix $M$ and from it the dual matrix $B$ using \S\ref{s-du} and Figure~\ref{f-1}. The biological context and the question being asked typically suggest one or more species of interest. For the initial subset of complexes, $Z$, choose all those complexes in which the species of interest have positive stoichiometry. Check if there are any invariants on $Z$ using Proposition~\ref{t-dim}. If not, then consider any additional complexes that have at least one species in common with the complexes in $Z$. Add each of these complexes to $Z$ in turn, starting with those that introduce the fewest new species and allowing the number of new species to increase as slowly as possible. With each addition, test for invariants as before. If this fails, consider adding the new complexes in groups, trying, as before, to minimise the number of new species that are introduced.

\begin{figure}
\centering 
\includegraphics[viewport=70 36 676 398,width=0.8\textwidth,height=\textheight,keepaspectratio]{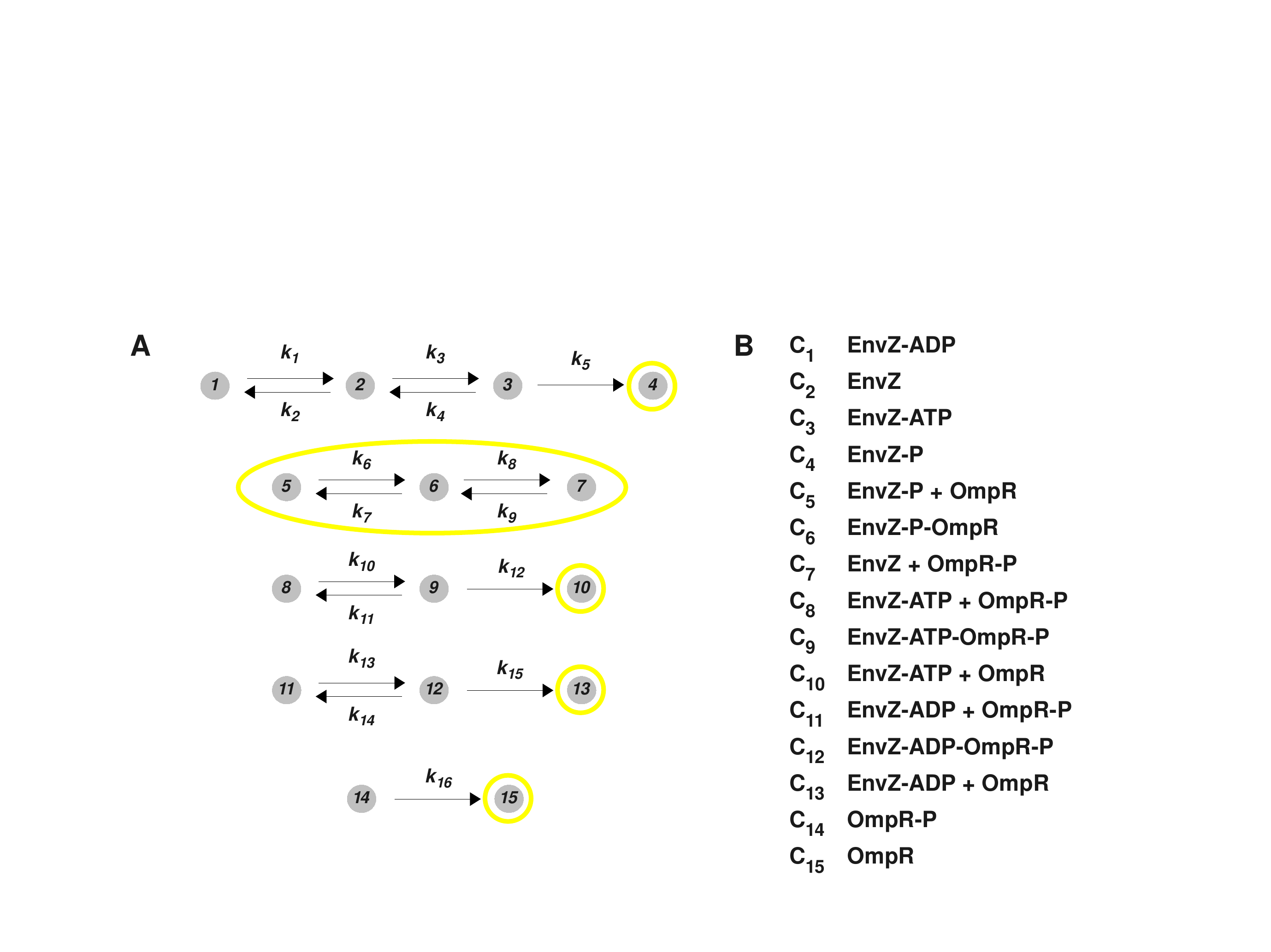}
\caption{Complexes for the modified EnvZ/OmpR network discussed in \S\ref{s-sys}, in which a single reaction is added to the network in Figure~\ref{f-1}B. {\bf A} Labelled, directed graph on the complexes, with the terminal SCCs outlined in yellow. Only the connected component at the bottom differs from the graph in Figure~\ref{f-1}C. {\bf B} Numbering scheme, with two additional complexes, $C_{14}$ and $C_{15}$, beyond those in Figure~\ref{f-1}D. \label{f-4}}
\end{figure}

To demonstrate this procedure, we use a modification of the EnvZ/OmpR network in Figure~\ref{f-2}B. We add a single new reaction
\[ \mbox{OmpR-P} \longrightarrow \mbox{OmpR} \]
for spontaneous (non-catalysed) dephosphorylation of OmpR-P. The phos\-pho-aspartate bond in response regulators is labile and may be spontaneously hydrolysed, so this new reaction is biochemically plausible. The labelled, directed graph for the modified network in Figure~\ref{f-4}A has fifteen nodes and sixteen edges. Each connected component still has only a single terminal SCC, like the graph in Figure~\ref{f-1}C, but now $\delta_D = \delta_S = 3$. The matrices $M$ and $B$ are provided in the SI, along with other details of the calculation. 

The biological context suggests that the active state of the response regulator, OmpR-P, is of most interest. Following the procedure above and using the table in Figure~\ref{f-4}B leads to $Z= \{C_7, C_8, C_{11}, C_{14}\}$ as an initial subset of complexes. It can be readily checked by inspection of $B$ that the corresponding rows yield a submatrix of full rank 4, so that Proposition~\ref{t-dim} tell us that there are no non-trivial invariants on $Z$. Among the remaining complexes, $C_1$, $C_2$ and $C_3$ each involve only species that are already present among the complexes in $Z$. Adding each to $Z$ in turn, it can be checked that each of the subsets $Z \cup \{C_1\}$, $Z \cup \{C_2\}$ and $Z \cup \{C_3\}$ have a space of invariants of dimension 1, which respectively yield the following non-trivial invariants, 
\begin{equation}
\begin{array}{c}
\textstyle k_{16}x^{C_{14}} - \left(\frac{k_1k_3k_5}{k_2(k_4+k_5)}\right)x^{C_1} + \left(\frac{k_{10}k_{12}}{k_{11}+k_{12}}\right)x^{C_8} + \left(\frac{k_{13}k_{15}}{k_{14}+k_{15}}\right)x^{C_{11}} = 0 \\[0.4em]
\textstyle k_{16}x^{C_{14}} - \left(\frac{k_3k_5}{k_4+k_5}\right)x^{C_2} + \left(\frac{k_{10}k_{12}}{k_{11}+k_{12}}\right)x^{C_8} + \left(\frac{k_{13}k_{15}}{k_{14}+k_{15}}\right)x^{C_{11}} = 0 \\[0.4em]
\textstyle k_{16}x^{C_{14}} - k_5x^{C_3} + \left(\frac{k_{10}k_{12}}{k_{11}+k_{12}}\right)x^{C_8} + \left(\frac{k_{13}k_{15}}{k_{14}+k_{15}}\right)x^{C_{11}} = 0 \,.
\end{array}
\label{e-3}
\end{equation}
These all have a similar form, due to the common subset $Z$. The absence of $C_7$ in these invariants could have been inferred directly from the pattern of entries in $B$ (SI).

A non-trivial invariant does not necessarily provide helpful biological insights. This depends crucially on the context and the question being studied. For instance, assuming that we have a positive steady-state, the second invariant in (\ref{e-3}) may be rewritten in terms of steady-state species concentrations as
\begin{equation}
[\mbox{OmpR-P}] = \frac{\left(\frac{k_3k_5}{k_4+k_5}\right)[\mbox{EnvZ}]}{k_{16} + \left(\frac{k_{10}k_{12}}{k_{11}+k_{12}}\right)[\mbox{EnvZ-ATP}] + \left(\frac{k_{13}k_{15}}{k_{14}+k_{15}}\right)[\mbox{EnvZ-ADP}]} \,.
\label{e-om}
\end{equation}
Equation (\ref{e-om}) establishes a steady-state relationship between the activated response regulator, OmpR-P, and the sensor kinase, EnvZ, under nucleotide loading. This may be useful depending on the available experimental data. It does suggest that OmpR-P no longer exhibits ACR, as it did for the network in Figure~\ref{f-1}B, since its steady-state level appears to depend on the amount of EnvZ present. However, not much more can be said just from (\ref{e-om}). 

The situation is different for the first and third invariants in (\ref{e-3}). They lead to a similar equation for [OmpR-P] as in (\ref{e-om}) but with the numerators on the right hand side given by, respectively,
\[ \left(\frac{k_1k_3k_5}{k_2(k_4+k_5)}\right)[\mbox{EnvZ-ADP}] ~~~\mbox{and}~~~ k_5[\mbox{EnvZ-ATP}] \,.\]
Because the same species now appears in both the numerator and the denominator, a simple comparison and cancellation, yields the inequalities
\begin{equation}
[\mbox{OmpR-P}] < \left\{\begin{array}{l}
                     \displaystyle \frac{k_1}{k_2}\frac{k_3k_5}{(k_4+k_5)}\frac{(k_{14}+k_{15})}{k_{13}k_{15}} \\[1em]
                     \displaystyle k_5\frac{(k_{11}+k_{12})}{k_{10}k_{12}} \,.
                     \end{array}\right.
\label{e-orp}
\end{equation}
The strictness of the inequality comes from the assumption that the steady state is positive. We see that the activated response regulator has two upper bounds that are robust: they depend only on the rate constants and not on the initial conditions or the total amounts of either EnvZ or OmpR. An interesting aspect of (\ref{e-orp}) is the absence of parameters $k_6, \cdots, k_9$, that relate to the phosphorylation of OmpR. The other reactions contribute factors to the bounds that can be biochemically interpreted. Recall that the catalytic efficiency of an enzyme is the ratio of its catalytic rate to its Michaelis-Menten constant, $k_{cat}/K_M$, \cite{cb95}. The factor $k_{13}k_{15}/(k_{14}+k_{15})$ in the first bound is the catalytic efficiency of OmpR-P dephosphorylation by EnvZ-ADP, while the factor $k_{10}k_{12}/(k_{11}+k_{12})$ in the second bound is the catalytic efficiency of OmpR-P dephosphorylation by EnvZ-ATP. The balance between these redundant dephosphorylation routes will influence which of the two bounds is the tighter and this balance is further modulated by the efficiency of nucleotide binding to EnvZ. The first bound is modulated by the factor $k_3k_5/(k_4+k_5)$, which may be treated as an effective catalytic efficiency for EnvZ phosphorylation by ATP (it has units of sec$^{-1}$ rather than M$^{-1}$sec$^{-1}$), and the factor $k_1/k_2$, which is the equilibrium constant for ADP binding. The second bound is only modulated by $k_5$, which is the catalytic rate for EnvZ phosphorylation. 

The existence of these robust bounds also strongly suggests that OmpR-P does not satisfy ACR. Indeed, it can be shown by algebraic calculation that [OmpR-P] can take different values in different steady states (SI). In contrast to the second invariant in (\ref{e-3}), the first and third invariants yield interesting and unexpected biological insights. Further non-trivial invariants can be sought using the procedure above and we leave this for the reader.

We note two further points of interest. First, the addition of a single new reaction to a network can markedly change its behaviour from exhibiting ACR to only having robust bounds. This is a feature of biochemical networks under mass-action kinetics. It raises difficult problems of interpretation because there is always the possibility that the actual cellular network may include reactions that have been missed in a model. Very little work has been done on this difficult question. Invariants may provide a way to study it: perhaps certain invariants can be shown to remain ``invariant'' when a network is enlarged in a particular way. Second, a theme is emerging from the examples considered here. Despite the differences in network structures between Figures~\ref{f-1}, \ref{f-2} and \ref{f-4}, the bifunctionality in each case serves to limit the steady state concentration of a substrate form, either absolutely, as in Figure~\ref{f-1}, or relative to some robust upper bound, as in Figures~\ref{f-2} and \ref{f-4}.  We speculate that this may be a design principle of those bifunctional enzymes that catalyse forward and reverse modifications. There are other forms of bifunctionality, such as enzymes that catalyse successive steps in a metabolic pathway, and preliminary studies suggest that these behave very differently. If modification bifunctionality did evolve to implement concentration control, markedly different network structures seem to have converged upon it.

\section{DISCUSSION}

The nonlinearity of molecular networks makes it impossible to solve their dynamical behaviour in closed form. Their analysis has therefore relied on numerical integration and simulation, for which the biochemical details and the numerical values of all parameters must be specified in advance. This has made it difficult, if not impossible, to ``see the wood for the trees'' and to discern general principles within the overwhelming molecular complexity of cellular processes. Invariants are part of a new approach in which network behaviour at steady state can be analysed with the parameters treated symbolically. There are now several examples, drawn from different biological contexts, in which the invariants summarise the essential steady-state properties of the network. The key to wide exploitation of this method is that it should be readily applicable to realistic networks. In principle, Gr\"obner bases allow any invariant to be calculated but this is computationally infeasible in practice. Complex-linear invariants form a limited subset of all invariants but, as shown here, they have biological significance and can be efficiently calculated for realistic networks. Our work clarifies previous results and provides a new tool for symbolic, steady-state analysis of molecular networks.




\section{ACKNOWLEDGEMENTS}

MPM and AD were partially supported by UBACYT 20020100100242, CONICET PIP 112-200801-00483, and ANPCyT PICT 2008-0902, Argentina. RLK, TD and JG were partially supported by NSF 0856285.

\section{Appendix}

The symbolic linear algebra calculations that follow may be readily undertaken in any computer algebra system. We used Mathematica, for which we wrote custom functions that compute the relevant matrices automatically from a description of the reaction network. 

\subsection{Invariants that are not of type 1}

Consider the hypothetical reaction network in Figure~\ref{f-1}A. While such chemistry is unlikely, it illustrates the mathematical issues. The network has three species, $S_1, S_2, S_3$ and nine complexes $C_1, \cdots, C_9$, ordered as in Figure~\ref{f-1}C. The ODEs are 
\begin{equation}
   \begin{array}{rcl}
   \displaystyle \frac{dx_1}{dt} & = & k_1x_1 \\[1em]
   \displaystyle \frac{dx_2}{dt} & = & k_4x_1x_3 - k_2x_2^2 \\[1em]
   \displaystyle \frac{dx_3}{dt} & = & k_5x_1x_2 - k_3x_3^2 \,.
   \end{array}
\label{e-ode}
\end{equation}
With the given ordering, the matrix $M = Y.\lap(G)$ is
\[\left( \begin{array}{ccccccccc}
   k_1 & 0 & 0 & 0 & 0 & 0 & 0 & 0 & 0 \\
   0 & 0 & -k_2 & 0 & 0 & 0 & k_4 & 0 & 0 \\
   0 & 0 & 0 & 0 & -k_3 & 0 & 0 & k_5 & 0
   \end{array}\right)\,.
\]
Focussing on the complexes $C_1, C_3, C_5, C_7, C_8$, Proposition~\ref{prop:1} shows that the space of type 1 complex-linear invariants has dimension three. However, it is easy to see from (\ref{e-ode}) that the only steady state of the network is when $x_1 = x_2 = x_3 = 0$. Hence, for any values of $a, b, c, d, e \in \R$, the polynomial expression
\[ ax^{C_1} + bx^{C_3} + cx^{C_5} + dx^{C_7} + ex^{C_8} \]
always vanishes in any steady state. Hence, the space of complex-linear invariants on $C_1, C_3, C_5, C_7, C_8$ has dimension five.
\begin{figure}
\centering 
\includegraphics[viewport=64 46 418 536,width=\textwidth,height=0.4\textheight,keepaspectratio]{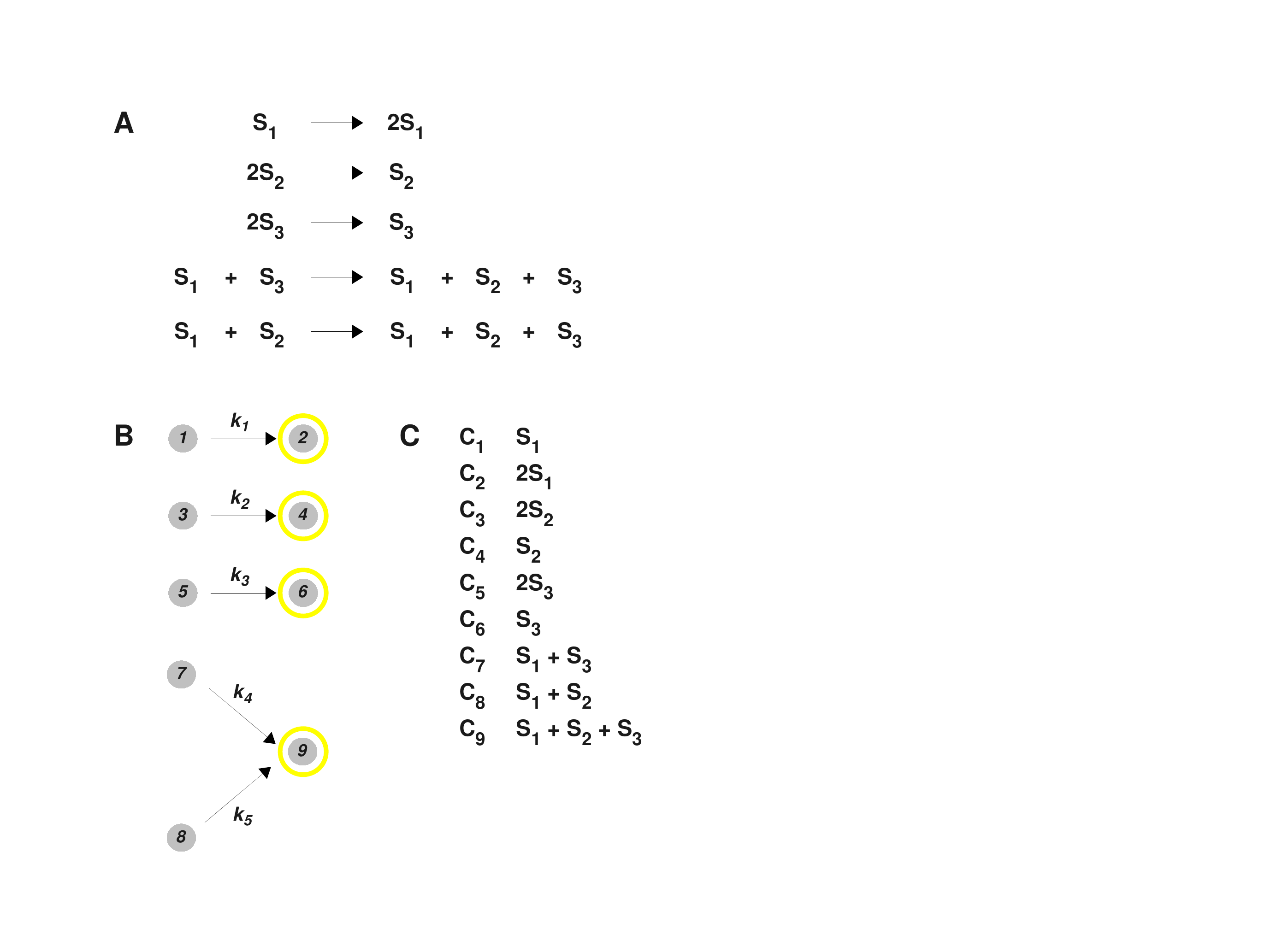}
\caption{Network with complex-linear invariants that are not of type 1. {\bf A} Hypothetical reaction network. {\bf B} Labelled, directed graph on the complexes, with the terminal strongly-connected components outlined in yellow. {\bf C} Numbering scheme for the complexes. \label{f-1}}
\end{figure}

\subsection{Corollary~\ref{cor:1}}

The nine species and thirteen complexes in the EnvZ/OmpR network in Figure~1 are ordered as follows.
\[
\begin{array}{|l|l||l|l|l|}
\hline
S_1 & \mbox{EnvZ-P-OmpR} & C_1 & S_8 & \mbox{EnvZ-ADP} \\
\hline
S_2 & \mbox{EnvZ-ATP-OmpR-P} & C_2 & S_4 & \mbox{EnvZ} \\
\hline
S_3 & \mbox{EnvZ-ADP-OmpR-P} & C_3 & S_7 & \mbox{EnvZ-ATP} \\
\hline
S_4 & \mbox{EnvZ} & C_4 & S_9 & \mbox{EnvZ-P} \\
\hline
S_5 & \mbox{OmpR} & C_5 & S_9 + S_5 & \mbox{EnvZ-P} + \mbox{OmpR} \\
\hline
S_6 & \mbox{OmpR-P} & C_6 & S_1 & \mbox{EnvZ-P-OmpR} \\
\hline
S_7 & \mbox{EnvZ-ATP} & C_7 & S_4 + S_6 & \mbox{EnvZ} + \mbox{OmpR-P} \\
\hline
S_8 & \mbox{EnvZ-ADP} & C_8 & S_7 + S_6 & \mbox{EnvZ-ATP} + \mbox{OmpR-P} \\
\hline
S_9 & \mbox{EnvZ-P} & C_9 & S_2 & \mbox{EnvZ-ATP-OmpR-P} \\
\hline
& & C_{10} & S_7 + S_5 & \mbox{EnvZ-ATP} + \mbox{OmpR} \\
\hline
& & C_{11} & S_8 + S_6 & \mbox{EnvZ-ADP} + \mbox{OmpR-P} \\
\hline
& & C_{12} & S_3 & \mbox{EnvZ-ADP-OmpR-P} \\
\hline
& & C_{13} & S_8 + S_5 & \mbox{EnvZ-ADP} + \mbox{OmpR} \\
\hline
\end{array}
\]
With this ordering and with the rate constants as in  Figure~1C, the matrix $M = Y.\lap(G)$ is
{\tiny
\[
\left(\begin{array}{ccccccccccccccc}
 0 & 0 & 0 & 0 & k_6 & -k_7-k_8 & k_9 & 0 & 0 & 0 & 0 & 0 & 0 \\
 0 & 0 & 0 & 0 & 0 & 0 & 0 & k_{10} & -k_{11}-k_{12} & 0 & 0 & 0 & 0 \\
 0 & 0 & 0 & 0 & 0 & 0 & 0 & 0 & 0 & 0 & k_{13} & -k_{14}-k_{15} & 0 \\
 k_1 & -k_2-k_3 & k_4 & 0 & 0 & k_8 & -k_9 & 0 & 0 & 0 & 0 & 0 & 0 \\
 0 & 0 & 0 & 0 & -k_6 & k_7 & 0 & 0 & k_{12} & 0 & 0 & k_{15} & 0 \\
 0 & 0 & 0 & 0 & 0 & k_8 & -k_9 & -k_{10} & k_{11} & 0 & -k_{13} & k_{14} & 0 \\
 0 & k_3 & -k_4-k_5 & 0 & 0 & 0 & 0 & -k_{10} & k_{11}+k_{12} & 0 & 0 & 0 & 0 \\
 -k_1 & k_2 & 0 & 0 & 0 & 0 & 0 & 0 & 0 & 0 & -k_{13} & k_{14}+k_{15} & 0 \\
 0 & 0 & k_5 & 0 & -k_6 & k_7 & 0 & 0 & 0 & 0 & 0 & 0 & 0
\end{array}\right)
\]
}
A basis for the kernel of $M$ can then be calculated to make up the columns of a matrix $B$.
\[
\begin{pmatrix}
 0 & \frac{k_2 \left(k_4+k_5\right) k_{15}}{k_1 k_3 k_5} & 0 & \frac{k_2 \left(k_4+k_5\right) k_{12}}{k_1 k_3 k_5} & 0 & 0 \\[0.3em]
 0 & \frac{\left(k_4+k_5\right) k_{15}}{k_3 k_5} & 0 & \frac{\left(k_4+k_5\right) k_{12}}{k_3 k_5} & 0 & 0 \\[0.3em]
 0 & \frac{k_{15}}{k_5} & 0 & \frac{k_{12}}{k_5} & 0 & 0 \\[0.3em]
 0 & 0 & 0 & 0 & 0 & 1 \\[0.3em]
 0 & \frac{\left(k_7+k_8\right) k_{15}}{k_6 k_8} & 0 & \frac{\left(k_7+k_8\right) k_{12}}{k_6 k_8} & \frac{k_7 k_9}{k_6 k_8} & 0 \\[0.3em]
 0 & \frac{k_{15}}{k_8} & 0 & \frac{k_{12}}{k_8} & \frac{k_9}{k_8} & 0 \\[0.3em]
 0 & 0 & 0 & 0 & 1 & 0 \\[0.3em]
 0 & 0 & 0 & \frac{k_{11}+k_{12}}{k_{10}} & 0 & 0 \\[0.3em]
 0 & 0 & 0 & 1 & 0 & 0 \\[0.3em]
 0 & 0 & 1 & 0 & 0 & 0 \\[0.3em]
 0 & \frac{k_{14}+k_{15}}{k_{13}} & 0 & 0 & 0 & 0 \\[0.3em]
 0 & 1 & 0 & 0 & 0 & 0 \\[0.3em]
 1 & 0 & 0 & 0 & 0 & 0
\end{pmatrix}
\]
Corollary~\ref{cor:1} focusses on the complexes $C_1, C_3, C_8, C_{11}$, so that $k = 4$. These are not the first four complexes in the ordering, as was assumed for convenience in \S\ref{ss:be}. We can imagine that the columns of $M$ and the rows of $B$ have been permuted so that these complexes are now the first in the ordering but we will not bother to write out these new matrices. We note that columns $2$ and $4$ of $B$ have non-zero entries in the relevant four rows, while the remaining columns have zero entries. We can undertake elementary column operations on $B$, as described in the paper (in fact, only interchange of columns is required), to bring $B$ into lower-triangular block form. The resulting $4 \times 2$ sub-matrix, $B'$, in Equation~\eqref{e-bf}, is then given by
\[
\begin{pmatrix}
 \frac{k_2 \left(k_4+k_5\right) k_{15}}{k_1 k_3 k_5} & \frac{k_2 \left(k_4+k_5\right) k_{12}}{k_1 k_3 k_5} \\[0.3em]
 \frac{k_{15}}{k_5} & \frac{k_{12}}{k_5} \\[0.3em]
 0 & \frac{k_{11}+k_{12}}{k_{10}} \\[0.3em]
 \frac{k_{14}+k_{15}}{k_{13}} & 0
\end{pmatrix}
\]
The columns of this are linearly independent, so that $\rk B' = 2$. It follows from Proposition~\ref{prop:1} that the dimension of the space of type 1 complex-linear invariants on $C_1, C_3, C_8, C_{11}$ is $2$, as claimed. To generate the invariants, we note that rows $2$ and $3$ of $B'$ are linearly independent, so that we can  take $U = \{2,3\}$ and $V = \{1,4\}$. Then
\[ 
B'_U = \begin{pmatrix}
       \frac{k_{15}}{k_5} & \frac{k_{12}}{k_5} \\[0.3em]
       0 & \frac{k_{11}+k_{12}}{k_{10}}
       \end{pmatrix}\,,\hspace{2em}
B'_V = \begin{pmatrix}
       \frac{k_2 \left(k_4+k_5\right) k_{15}}{k_1 k_3 k_5} & \frac{k_2 \left(k_4+k_5\right) k_{12}}{k_1 k_3 k_5} \\[0.3em]
       \frac{k_{14}+k_{15}}{k_{13}} & 0
       \end{pmatrix}
\]
Since $\Psi(x)_U = (x^{C_3},x^{C_8})^\dagger$ and $\Psi(x)_V = (x^{C_1},x^{C_{11}})^\dagger$, the two linearly independent type 1 complex-linear invariants may be read off from Equation~\eqref{e-inv},
\[ 
   \begin{pmatrix}
   x^{C_1} \\
   x^{C_{11}}
   \end{pmatrix} =
   \begin{pmatrix}
   \frac{k_2(k_4 + k_5)}{k_1k_3} & 0 \\[0.3em]
   \frac{k_5(k_{14}+k_{15})}{k_{13}k_{15}} & -\frac{k_{10}k_{12}(k_{14}+k_{15})}{k_{13}k_{15}(k_{11}+k_{12})}
   \end{pmatrix}
   \begin{pmatrix}
   x^{C_3} \\
   x^{C_8}
   \end{pmatrix}
\]
to yield the expressions in Paper Corollary~\ref{cor:1}, as claimed.

\subsection{Corollary~\ref{cor:2}}

The eight species and fourteen complexes of the PFK-2/FBPase-2 network in Paper Figures~2 and 3 are ordered as follows.
\[
\begin{array}{|l|l||l|l|l|}
\hline
S_1 & \mbox{F2,6BP} & C_1 & S_5 & \mbox{E} \\
\hline
S_2 & \mbox{F6P} & C_2 & S_7 & \mbox{E-ATP} \\
\hline
S_3 & \mbox{E-ATP-F6P} & C_3 & S_7 + S_2 & \mbox{E-ATP} + \mbox{F6P} \\
\hline
S_4 & \mbox{E-ATP-F6P-F2,6BP} & C_4 & S_3 & \mbox{E-ATP-F6P} \\
\hline
S_5 & \mbox{E} & C_5 & S_5 + S_1 & \mbox{E} + \mbox{F2,6BP} \\
\hline
S_6 & \mbox{E-F2,6BP} & C_6 & S_6 & \mbox{E-F2,6BP} \\
\hline
S_7 & \mbox{E-ATP} & C_7 & S_5 + S_2 & \mbox{E} + \mbox{F6P} \\
\hline
S_8 & \mbox{E-ATP-F2,6BP} & C_8 & S_8  & \mbox{E-ATP-F2,6BP} \\
\hline
& & C_9 & S_7 + S_1 & \mbox{E-ATP} + \mbox{F2,6BP} \\
\hline
& & C_{10} & S_3 + S_1 & \mbox{E-ATP-F6P} + \mbox{F2,6BP} \\
\hline
& & C_{11} & S_4 & \mbox{E-ATP-F6P-F2,6BP} \\
\hline
& & C_{12} & S_8 + S_2 & \mbox{E-ATP-F2,6BP} + \mbox{F6P} \\
\hline
& & C_{13} & S_6 + S_1 & \mbox{E-F2,6BP} + \mbox{F2,6BP} \\
\hline
& & C_{14} & S_3 + S_2 & \mbox{E-ATP-F6P} + \mbox{F6P} \\
\hline
\end{array}
\]
With this ordering and with the rate constants in Paper Figure~3A, the matrix $M = Y.\lap(G)$ is
{\tiny
\[
\left(
\begin{array}{cccccccccccccc}
0 & 0 & 0 & k_5 & -k_6 & k_7 & 0 & k_{13} & -k_{12} & -k_{14} & k_{15}+k_{18} & 0 & 0 & 0 \\
 0 & 0 & -k_3 & k_4 & 0 & k_8 & 0 & k_{11} & 0 & 0 & k_{17}+k_{19} & -k_{16} & 0 & 0 \\
 0 & 0 & k_3 & -k_4-k_5 & 0 & 0 & 0 & 0 & 0 & -k_{14} & k_{15}+k_{19} & 0 & 0 & 0 \\
 0 & 0 & 0 & 0 & 0 & 0 & 0 & 0 & 0 & k_{14} & -k_{15}-k_{17}-k_{18}-k_{19} & k_{16} & 0 & 0 \\
 -k_1 & k_2 & 0 & k_5 & -k_6 & k_7+k_8 & 0 & 0 & 0 & 0 & 0 & 0 & 0 & 0 \\
 0 & 0 & 0 & 0 & k_6 & -k_7-k_8-k_{10} & 0 & k_9 & 0 & 0 & k_{18} & 0 & 0 & 0 \\
 k_1 & -k_2 & -k_3 & k_4 & 0 & 0 & 0 & k_{11}+k_{13} & -k_{12} & 0 & 0 & 0 & 0 & 0 \\
 0 & 0 & 0 & 0 & 0 & k_{10} & 0 & -k_9-k_{11}-k_{13} & k_{12} & 0 & k_{17} & -k_{16} & 0 & 0
\end{array}
\right)
\]
}
and a matrix $B$, whose columns form a basis for the kernel of $M$, can then be calculated as
{\tiny
\[
\begin{pmatrix}
0 & 0 & \frac{k_8-k_{10}}{k_1 k_{10}} & \frac{k_{19}}{k_1} & \frac{\left(-k_8+k_{10}\right) k_{12}}{k_1 k_{10}} & \frac{k_8 k_9+k_{10} k_{11}}{k_1
k_{10}} & 0 & \frac{k_2}{k_1} \\[0.4em]
 0 & 0 & 0 & 0 & 0 & 0 & 0 & 1 \\[0.4em]
 0 & 0 & \frac{-k_5k_{10}+(k_4+k_5)k_8}{k_3k_5 k_{10}} & \frac{-k_4 k_{18}+\left(k_4+k_5\right) k_{19}}{k_3 k_5} & -\frac{\left(k_4+k_5\right) k_8 k_{12}}{k_3 k_5 k_{10}} & \frac{\left(k_4+k_5\right) \left(k_8 k_9+k_{10} k_{11}\right)}{k_3 k_5 k_{10}} & 0 & 0 \\[0.4em]
 0 & 0 & \frac{k_8}{k_5 k_{10}} & \frac{-k_{18}+k_{19}}{k_5} & -\frac{k_8 k_{12}}{k_5 k_{10}} & \frac{k_8 k_9+k_{10} k_{11}}{k_5 k_{10}} & 0 & 0 \\[0.4em]
 0 & 0 & \frac{k_7+k_8+k_{10}}{k_6 k_{10}} & -\frac{k_{18}}{k_6} & -\frac{\left(k_7+k_8+k_{10}\right) k_{12}}{k_6 k_{10}} & \frac{\left(k_7+k_8\right)k_9}{k_6 k_{10}} & 0 & 0 \\[0.4em]
 0 & 0 & \frac{1}{k_{10}} & 0 & -\frac{k_{12}}{k_{10}} & \frac{k_9}{k_{10}} & 0 & 0 \\[0.4em]
 0 & 0 & 0 & 0 & 0 & 0 & 1 & 0 \\[0.4em]
 0 & 0 & 0 & 0 & 0 & 1 & 0 & 0 \\[0.4em]
 0 & 0 & 0 & 0 & 1 & \frac{k_{11}+k_{13}}{k_{12}} & 0 & 0 \\[0.4em]
 0 & 0 & -\frac{1}{k_{14}} & \frac{k_{15}+k_{18}+k_{19}}{k_{14}} & 0 & 0 & 0 & 0 \\[0.4em]
 0 & 0 & 0 & 1 & 0 & 0 & 0 & 0 \\[0.4em]
 0 & 0 & \frac{1}{k_{16}} & \frac{k_{17}}{k_{16}} & 0 & 0 & 0 & 0 \\[0.4em]
 0 & 1 & 0 & 0 & 0 & 0 & 0 & 0 \\[0.4em]
 1 & 0 & 0 & 0 & 0 & 0 & 0 & 0
\end{pmatrix}
\]
}
Corollary~\ref{cor:2} focusses on the complexes $C_1, C_2, C_4, C_6, C_8, C_{11}$, so that $k = 6$. As before, we can imagine that the columns of $M$ and the rows of $B$ have been permuted to make these complexes first in the ordering. Only columns $3, 4, 5, 6, 8$ of $B$ have non-zero entries in the relevant rows and, when restricted to these rows, column $5$ is a scalar multiple of column $3$. As before, we can interchange columns to bring $B$ into lower-triangular block form, with the resulting $6 \times 4$ sub-matrix, $B'$, in Equation~\eqref{e-bf} given by
\[
\begin{pmatrix}
\frac{k_8-k_{10}}{k_1 k_{10}} & \frac{k_{19}}{k_1} & \frac{k_8 k_9+k_{10} k_{11}}{k_1 k_{10}} & \frac{k_2}{k_1} \\[0.3em]
 0 & 0 & 0 & 1 \\[0.3em]
 \frac{k_8}{k_5 k_{10}} & \frac{-k_{18}+k_{19}}{k_5} & \frac{k_8 k_9+k_{10} k_{11}}{k_5 k_{10}} & 0 \\[0.3em]
 \frac{1}{k_{10}} & 0 & \frac{k_9}{k_{10}} & 0 \\[0.3em]
 0 & 0 & 1 & 0 \\[0.3em]
 0 & 1 & 0 & 0
\end{pmatrix}
\]
with $B'$ evidently of full rank $4$. It follows from Paper Proposition~1 that the space of type 1 complex-linear invariants on $C_1, C_2, C_4, C_6, C_8, C_{11}$ has dimension two, as claimed. To generate the invariants, we can take $U = \{2, 4, 5, 6\}$ and $V = \{1, 3\}$ so that 
\[ 
B'_U = \begin{pmatrix}
 0 & 0 & 0 & 1 \\[0.3em]
 \frac{1}{k_{10}} & 0 & \frac{k_9}{k_{10}} & 0 \\[0.3em]
 0 & 0 & 1 & 0 \\[0.3em]
 0 & 1 & 0 & 0
\end{pmatrix}\,,\hspace{1em}
B'_V = \begin{pmatrix}
\frac{k_8-k_{10}}{k_1 k_{10}} & \frac{k_{19}}{k_1} & \frac{k_8 k_9+k_{10} k_{11}}{k_1 k_{10}} & \frac{k_2}{k_1} \\[0.3em]
 \frac{k_8}{k_5 k_{10}} & \frac{-k_{18}+k_{19}}{k_5} & \frac{k_8 k_9+k_{10} k_{11}}{k_5 k_{10}} & 0
\end{pmatrix}
\]
$B'_U$ is evidently non-singular. The two linear-independent type 1 complex-linear invariants can then be read off from Equation~\eqref{e-inv},
\[ 
   \begin{pmatrix}
   x^{C_1} \\
   x^{C_4}
   \end{pmatrix} =
   \begin{pmatrix}
   \frac{k_2}{k_1} & \frac{(k_8-k_{10})}{k_1} & \frac{(k_9+k_{11})}{k_1} & \frac{k_{19}}{k_1} \\[0.3em]
   0 & \frac{k_8}{k_5} & \frac{k_{11}}{k_5} & \frac{(-k_{18} + k_{19})}{k_5} \\[0.3em]
   \end{pmatrix}
   \begin{pmatrix}
   x^{C_2} \\
   x^{C_6} \\
   x^{C_8} \\
   x^{C_{11}}
   \end{pmatrix}
\]
to yield the expressions in Paper Corollary~\ref{cor:2}, as claimed.

\subsection{Example of the systematic procedure}
\label{s-sp}

The example in \S~\ref{s-sys} is a modification of that in Corollary~\ref{cor:1}. As shown in Figure~5 it has nine species and fifteen complexes which are ordered as follows.
\[
\begin{array}{|l|l||l|l|l|}
\hline
S_1 & \mbox{EnvZ-P-OmpR} & C_1 & S_8 & \mbox{EnvZ-ADP} \\
\hline
S_2 & \mbox{EnvZ-ATP-OmpR-P} & C_2 & S_4 & \mbox{EnvZ} \\
\hline
S_3 & \mbox{EnvZ-ADP-OmpR-P} & C_3 & S_7 & \mbox{EnvZ-ATP} \\
\hline
S_4 & \mbox{EnvZ} & C_4 & S_9 & \mbox{EnvZ-P} \\
\hline
S_5 & \mbox{OmpR} & C_5 & S_9 + S_5 & \mbox{EnvZ-P} + \mbox{OmpR} \\
\hline
S_6 & \mbox{OmpR-P} & C_6 & S_1 & \mbox{EnvZ-P-OmpR} \\
\hline
S_7 & \mbox{EnvZ-ATP} & C_7 & S_4 + S_6 & \mbox{EnvZ} + \mbox{OmpR-P} \\
\hline
S_8 & \mbox{EnvZ-ADP} & C_8 & S_7 + S_6 & \mbox{EnvZ-ATP} + \mbox{OmpR-P} \\
\hline
S_9 & \mbox{EnvZ-P} & C_9 & S_2 & \mbox{EnvZ-ATP-OmpR-P} \\
\hline
& & C_{10} & S_7 + S_5 & \mbox{EnvZ-ATP} + \mbox{OmpR} \\
\hline
& & C_{11} & S_8 + S_6 & \mbox{EnvZ-ADP} + \mbox{OmpR-P} \\
\hline
& & C_{12} & S_3 & \mbox{EnvZ-ADP-OmpR-P} \\
\hline
& & C_{13} & S_8 + S_5 & \mbox{EnvZ-ADP} + \mbox{OmpR} \\
\hline
& & C_{14} & S_6 & \mbox{OmpR-P} \\
\hline
& & C_{15} & S_5 & \mbox{OmpR} \\
\hline
\end{array}
\]
With this ordering and with the rate constants shown in Figure~5A, the matrix $M$ extends that for Corollary~\ref{cor:1} with a $9 \times 2$ block on the right:
{\small
\[
\scriptsize
\left(
\begin{array}{ccccccccccccccc}
 0 & 0 & 0 & 0 & k_6 & -k_7-k_8 & k_9 & 0 & 0 & 0 & 0 & 0 & 0 & 0 & 0 \\
 0 & 0 & 0 & 0 & 0 & 0 & 0 & k_{10} & -k_{11}-k_{12} & 0 & 0 & 0 & 0 & 0 & 0 \\
 0 & 0 & 0 & 0 & 0 & 0 & 0 & 0 & 0 & 0 & k_{13} & -k_{14}-k_{15} & 0 & 0 & 0 \\
 k_1 & -k_2-k_3 & k_4 & 0 & 0 & k_8 & -k_9 & 0 & 0 & 0 & 0 & 0 & 0 & 0 & 0 \\
 0 & 0 & 0 & 0 & -k_6 & k_7 & 0 & 0 & k_{12} & 0 & 0 & k_{15} & 0 & k_{16} & 0 \\
 0 & 0 & 0 & 0 & 0 & k_8 & -k_9 & -k_{10} & k_{11} & 0 & -k_{13} & k_{14} & 0 & -k_{16} & 0 \\
 0 & k_3 & -k_4-k_5 & 0 & 0 & 0 & 0 & -k_{10} & k_{11}+k_{12} & 0 & 0 & 0 & 0 & 0 & 0 \\
 -k_1 & k_2 & 0 & 0 & 0 & 0 & 0 & 0 & 0 & 0 & -k_{13} & k_{14}+k_{15} & 0 & 0 & 0 \\
 0 & 0 & k_5 & 0 & -k_6 & k_7 & 0 & 0 & 0 & 0 & 0 & 0 & 0 & 0 & 0
\end{array}
\right) \,.
\]
}
A matrix $B$, whose columns form a basis for $\ker M$, has the following form.
\[
B = \begin{pmatrix}
 0 & \frac{k_2 \left(k_4+k_5\right) k_{16}}{k_1 k_3 k_5} & 0 & \frac{k_2 \left(k_4+k_5\right) k_{15}}{k_1 k_3 k_5} & 0 & \frac{k_2 \left(k_4+k_5\right) k_{12}}{k_1 k_3 k_5} & 0 & 0 \\[0.3em]
 0 & \frac{\left(k_4+k_5\right) k_{16}}{k_3 k_5} & 0 & \frac{\left(k_4+k_5\right) k_{15}}{k_3 k_5} & 0 & \frac{\left(k_4+k_5\right) k_{12}}{k_3 k_5} & 0 & 0 \\[0.3em]
 0 & \frac{k_{16}}{k_5} & 0 & \frac{k_{15}}{k_5} & 0 & \frac{k_{12}}{k_5} & 0 & 0 \\[0.3em]
 0 & 0 & 0 & 0 & 0 & 0 & 0 & 1 \\[0.3em]
 0 & \frac{\left(k_7+k_8\right) k_{16}}{k_6 k_8} & 0 & \frac{\left(k_7+k_8\right) k_{15}}{k_6 k_8} & 0 & \frac{\left(k_7+k_8\right) k_{12}}{k_6 k_8} & \frac{k_7 k_9}{k_6 k_8} & 0 \\[0.3em]
 0 & \frac{k_{16}}{k_8} & 0 & \frac{k_{15}}{k_8} & 0 & \frac{k_{12}}{k_8} & \frac{k_9}{k_8} & 0 \\[0.3em]
 0 & 0 & 0 & 0 & 0 & 0 & 1 & 0 \\[0.3em]
 0 & 0 & 0 & 0 & 0 & \frac{k_{11}+k_{12}}{k_{10}} & 0 & 0 \\[0.3em]
 0 & 0 & 0 & 0 & 0 & 1 & 0 & 0 \\[0.3em]
 0 & 0 & 0 & 0 & 1 & 0 & 0 & 0 \\[0.3em]
 0 & 0 & 0 & \frac{k_{14}+k_{15}}{k_{13}} & 0 & 0 & 0 & 0 \\[0.3em]
 0 & 0 & 0 & 1 & 0 & 0 & 0 & 0 \\[0.3em]
 0 & 0 & 1 & 0 & 0 & 0 & 0 & 0 \\[0.3em]
 0 & 1 & 0 & 0 & 0 & 0 & 0 & 0 \\[0.3em]
 1 & 0 & 0 & 0 & 0 & 0 & 0 & 0
\end{pmatrix}\,.
\]
Consider the initial set of complexes, $Z = \{C_7, C_8, C_{11}, C_{14}\}$, suggested by the procedure in the Paper. The corresponding rows of $B$ have only a single non-zero entry, each of which occurs in a distinct column, so that the four rows constitute a sub-matrix of full rank. It follows from Proposition~\ref{prop:1} that the space of type 1 complex-linear invariants is empty. Next consider adding in turn each of the complexes $C_1$, $C_2$ and $C_3$ to $Z$. Consider first $Z \cup \{C_1\}$. The rows of $B$ may be permuted so that the rows corresponding to complexes $C_1, C_8, C_{11}, C_{14}, C_7$ are placed first and in that order. Permuting the columns of $B$ so that the columns $2, 4, 6, 7$ are placed first yields a matrix in lower-triangular block form, with the top left block, $B'$, given by
\[
B' = \left(
\begin{array}{cccc}
 \frac{k_2 \left(k_4+k_5\right) k_{16}}{k_1 k_3 k_5} & \frac{k_2 \left(k_4+k_5\right) k_{15}}{k_1 k_3 k_5} & \frac{k_2 \left(k_4+k_5\right) k_{12}}{k_1 k_3 k_5} & 0 \\[0.3em]
 0 & 0 & \frac{k_{11}+k_{12}}{k_{10}} & 0 \\[0.3em]
 0 & \frac{k_{14}+k_{15}}{k_{13}} & 0 & 0\\[0.3em]
 1 & 0 & 0 & 0 \\[0.3em]
 0 & 0 & 0 & 1 
\end{array}
\right)\,.
\]
It is evident that $B'$ is in block diagonal form, with the last row and column, corresponding to complex $C_7$, forming an identify matrix. It is then easy to see from Equation~\eqref{e-inv} that $C_7$ cannot appear in any invariant. In other words, it is sufficient to work with $C_8, C_{11}, C_{14}$, when adding $C_1$. Before doing this, note that the $C_1$, $C_2$ and $C_3$ rows of $B$ may be written in the form
\[ \alpha_i (0 ~~~ k_{16} ~~~ 0 ~~~ k_{15} ~~~ 0 ~~~ k_{12} ~~~ 0 ~~~ 0) \]
where
\begin{equation}
\alpha_1 = \frac{k_2(k_4+k_5)}{k_1k_3k_5} ~,~~~ \alpha_2 = \frac{(k_4+k_5)}{k_3k_5} ~,~~~ \alpha_3 = \frac{1}{k_5} \,.
\label{e-ai}
\end{equation}
Accordingly, we can do all three calculations at once by omitting $C_7$ and re-writing $B'$ in the form
\[
B' = \left(
\begin{array}{ccc}
 \alpha_ik_{16} & \alpha_ik_{15} & \alpha_ik_{12} \\[0.3em]
 0 & 0 & \frac{k_{11}+k_{12}}{k_{10}} \\[0.3em]
 0 & \frac{k_{14}+k_{15}}{k_{13}} & 0 \\[0.3em]
 1 & 0 & 0
\end{array}
\right) \,,
\]
with $\alpha_i$ given by (\ref{e-ai}) for $i = 1, 2, 3$. It is evident that $\rk B'= 3$ and we can choose $U = \{1,2,3\}$ and $V = \{4\}$, so that
\[
B'_U =
\left(
\begin{array}{ccc}
 \alpha_ik_{16} & \alpha_ik_{15} & \alpha_ik_{12} \\[0.3em]
 0 & 0 & \frac{k_{11}+k_{12}}{k_{10}} \\[0.3em]
 0 & \frac{k_{14}+k_{15}}{k_{13}} & 0
\end{array}
\right)\,, \hspace{2em}
B'_V = 
(1 ~~~ 0 ~~~ 0)\,.
\]
It follows that 
\[
(B'_U)^{-1} = 
\left(
\begin{array}{ccc}
 \frac{1}{\alpha_ik_{16}} & -\frac{k_{10} k_{12}}{\left(k_{11}+k_{12}\right) k_{16}} & -\frac{k_{13} k_{15}}{\left(k_{14}+k_{15}\right) k_{16}} \\[0.3em]
 0 & 0 & \frac{k_{13}}{k_{14}+k_{15}} \\[0.3em]
 0 & \frac{k_{10}}{k_{11}+k_{12}} & 0
\end{array}
\right).
\]
Note that $\alpha_i$ only appears in a single entry. Using Paper Equation (6), with $\Psi(x)_U = (x^{C_i},x^{C_8},x^{C_{11}})^\dagger$ for $i = 1, 2, 3$, and $\Psi(x)_V = (x^{C_{14}})$, we recover the invariants in Equation~\eqref{e-3}.

\subsection{Failure of ACR for the example in \S\ref{s-sys}}

The steady states of the example just discussed can be algebraically simplified as follows. The differential equations governing the system can be obtained from the matrix $M$ in \S\ref{s-sp} above by using the fundamental decomposition of CRNT, $dx/dt = M.\Psi(x)$. With the notation in the Table in \S\ref{s-sp}, this gives
{\small
\begin{eqnarray}
\frac{dS_1}{dt} & = & k_6S_5S_9 - (k_7+k_8)S_1 + k_9S_4S_6 \label{e-f1} \\ 
\frac{dS_2}{dt} & = & k_{10}S_6S_7 - (k_{11} + k_{12})S_2 \label{e-f2} \\ 
\frac{dS_3}{dt} & = & k_{13}S_6S_8 - (k_{14} + k_{15})S_3 \label{e-f3} \\ 
\frac{dS_4}{dt} & = & k_1S_8 - (k_2+k_3)S_4 + k_4S_7 + k_8S_1 - k_9S_4S_6 \label{e-f4} \\ 
\frac{dS_5}{dt} & = & k_7S_1 - k_6S_5S_9 + k_{12}S_2 + k_{15}S_3 + k_{16}S_6 \label{e-f5} \\ 
\frac{dS_6}{dt} & = & k_8S_1 - k_9S_4S_6 - k_{10}S_6S_7 + k_{11}S_2 - k_{13}S_6S_8 + k_{14}S_3 - k_{16}S_6 \label{e-f6} \\ 
\frac{dS_7}{dt} & = & k_3S_4 - (k_4+k_5)S_7 - k_{10}S_6S_7 + (k_{11}+k_{12})S_2 \label{e-f7} \\ 
\frac{dS_8}{dt} & = & k_2S_4 - k_1S_8 - k_{13}S_6S_8 + (k_{14}+k_{15})S_3 \label{e-f8} \\
\frac{dS_9}{dt} & = & k_5S_7 - k_6S_5S_9 + k_7S_1 \,. \label{e-f9}
\end{eqnarray}
}
It can be checked that these equations (or, equivalently, the matrix $M$ from which they are derived) satisfy two conservation laws, corresponding to the total amounts of sensor and response regulator, 
\begin{equation}
\begin{array}{rcl}
S_1 + S_2 + S_3 + S_4 + S_7 + S_8 + S_9 & = & K_1 \\
S_1 + S_2 + S_3 + S_5 + S_6 & = & K_2 \,,
\end{array}
\label{e-c12}
\end{equation}
where $K_1, K_2 \in \Rmas$ are constants determined by the initial conditions.

The steady states are obtained by setting the right hand sides of equations (\ref{e-f1})-(\ref{e-f9}) to zero. The nine variables may be partitioned into two subsets, $\{S_1, S_2, S_3, S_7, S_8\}$ and $\{S_4, S_5, S_6, S_9\}$, in such a way that the variables in the first subset can be written in terms of the variables in the second subset. Using equation (\ref{e-f1}),
\begin{equation}
S_1 = \left(\frac{k_6}{k_7 + k_8}\right)S_5S_9 + \left(\frac{k_9}{k_7 + k_8}\right)S_4S_6 \,.
\label{e-g1}
\end{equation}
Combining equations (\ref{e-f2}) and (\ref{e-f7}), we get
\begin{equation}
S_7 = \left(\frac{k_3}{k_4 + k_5}\right)S_4
\label{e-g7} 
\end{equation}
and, similarly, combining equations (\ref{e-f3}) and (\ref{e-f8}) we get
\begin{equation}
S_8 = \left(\frac{k_2}{k_1}\right)S_4 \,.
\label{e-g8} 
\end{equation}
Using equation (\ref{e-f2}) together with (\ref{e-g7}) we get
\begin{equation}
S_2 = \left(\frac{k_{10}}{k_{11}+k_{12}}\right)\left(\frac{k_3}{k_4 + k_5}\right)S_4S_6
\label{e-g2} 
\end{equation}
and, similarly, using equations (\ref{e-f3}) and (\ref{e-g8}) we get
\begin{equation}
S_3 = \left(\frac{k_{13}}{k_{14}+k_{15}}\right)\left(\frac{k_2}{k_1}\right)S_4S_6 \,.
\label{e-g3} 
\end{equation}
Equations (\ref{e-g1})-(\ref{e-g3}) describe $\{S_1, S_2, S_3, S_7, S_8\}$ in terms of $\{S_4, S_5, S_6, S_9\}$. 

If we now substitute in equation (\ref{e-f5}) for $S_1$, $S_2$ and $S_3$ using equations (\ref{e-g1}), (\ref{e-g2}) and (\ref{e-g3}), respectively, and simplify, we get
\begin{equation}
(\alpha S_4 + k_{16})S_6 = \left(\frac{k_6k_8}{k_7+k_8}\right)S_5S_9 \,
\label{e-u1}
\end{equation}
where $\alpha$ depends only on the rate constants and can be conveniently represented as $\alpha = \beta + \gamma$, where
\[ \beta = \left(\frac{k_7k_9}{k_7+k_8}\right) \,,\hspace{1em} \gamma = \left(\frac{k_{10}k_{12}}{k_{11}+k_{12}}\right)\left(\frac{k_3}{k_4 + k_5}\right) + \left(\frac{k_{13}k_{15}}{k_{14}+k_{15}}\right)\left(\frac{k_2}{k_1}\right) \,. \]
If we also substitute in equation (\ref{e-f9}) for $S_1$ and $S_7$ using equations (\ref{e-g1}) and (\ref{e-g7}), respectively, and simplify, we get
\begin{equation}
\left(\frac{k_3k_5}{k_4+k_5}\right)S_4 + \beta S_4S_6 = \left(\frac{k_6k_8}{k_7+k_8}\right)S_5S_9 \,.
\label{e-u2}
\end{equation}
Combining equations (\ref{e-u1}) and (\ref{e-u2}) we can express $S_6$ as a rational function of $S_4$,
\begin{equation}
S_6 = \frac{k_3k_5S_4}{(k_4+k_5)(\gamma S_4 + k_{16})} \,.
\label{e-g6}
\end{equation}
Finally, substituting this expression into equation (\ref{e-u1}), we can express $S_5$ as a rational function of $S_4$ and $S_9$, 
\begin{equation}
S_5 = \frac{k_3k_5(k_7+k_8)(\alpha S_4 + k_{16})S_4}{k_6k_8(k_4+k_5)(\gamma S_4 + k_{16})S_9} \,.
\label{e-g5}
\end{equation}
If $S_4$ and $S_9$ are given arbitrary positive values, the values of all the other variables are determined by equations (\ref{e-g5}), (\ref{e-g6}) and (\ref{e-g1}) to (\ref{e-g3}). It can be checked that these values form a positive steady-state of the system. The free quantities $S_4$ and $S_9$, in terms of which the other variables have been parameterised, implicitly determine the values of the conserved quantities $K_1$ and $K_2$ in equation (\ref{e-c12}). It can be seen from equation (\ref{e-g6}) that $S_6$, which is the concentration of the activated response regulator, $S_6 = [\mbox{OmpR-P}]$, varies with the choice of $S_4$ and, hence, does not exhibit ACR. 

It also follows from equation (\ref{e-g6}) that
\begin{equation}
[\mbox{OmpR-P}] = S_6 < \frac{k_3k_5}{(k_4+k_5)\gamma} \,,
\label{e-bd}
\end{equation}
which shows, as deduced in \S\ref{s-sys}, that OmpR-P has a robust upper bound. The bound in (\ref{e-bd}) is half the harmonic mean of the robust bounds in Equation~\ref{e-orp} and is therefore tighter than either of the latter. This is evident from equation (\ref{e-g6}), where it is clear that $S_6$ asymptotically approaches the bound in (\ref{e-bd}) as $S_4$ increases. Hence, (\ref{e-bd}) is the best possible bound on $S_6$. 


\begin{thebibliography}{22}

\bibitem[{BG03}]{bg03}
Batchelor, E., Goulian, M., 2003. Robustness and the cycle of phosphorylation
  and dephosphorylation in a two-component regulatory system. Proc. Natl. Acad.
  Sci. USA 100, 691--6.

\bibitem[{C-B95}]{cb95}
Cornish-Bowden, A., 1995. Fundamentals of Enzyme Kinetics, 2nd Edition.
  Portland Press, London, UK.

\bibitem[{CLO97}]{clos}
Cox, D., Little, J., O'Shea, D., 1997. Ideals, Varieties and Algorithms, 2nd
  Edition. Springer.

\bibitem[{CDSS08}]{cdss08}
Craciun, G., Dickenstein, A., Shiu, A., Sturmfels, B., 2009. Toric dynamical
  systems. J. Symb. Comp. 44, 1551--65.

\bibitem[{DCHLADG10}]{dcg10}
Dasgupta, T., Croll, D.~H., Heiden, M. H.~V., Locasale, J.~W., Alon, U.,
  Cantley, L.~C., Gunawardena, J., 2010. Bifunctionality in {PFK2/F2,6BPase}
  confers both robustness and plasticity in the control of glycolysis,
  submitted.

\bibitem[{F79}]{fein79}
Feinberg, M., 1979. Lectures on {C}hemical {R}eaction {N}etworks, lecture
  notes, {M}athematics {R}esearch {Center}, {U}niversity of {W}isconsin.

\bibitem[{FH77}]{fh77}
Feinberg, M., Horn, F., 1977. Chemical mechanism structure and the coincidence
  of the stoichiometric and kinetic subspace. Arch. Rational Mech. Anal. 66,
  83--97.

\bibitem[{G03}]{gun03}
Gunawardena, J., 2003. Chemical {R}eaction {N}etwork {T}heory for {\em
  in-silico} biologists, {L}ecture notes, {H}arvard {U}niv, 2003.
  \verb|vcp.med.harvard.edu/papers/crnt.pdf|.

\bibitem[{G10}]{gun-misb-1}
Gunawardena, J., 2010. Models in systems biology: the parameter problem and the
  meanings of robustness. In: Lodhi, H., Muggleton, S. (Eds.), Elements of
  Computational Systems Biology. Wiley Book Series on Bioinformatics. John
  Wiley and Sons, Inc.

\bibitem[{G12}]{gun-mt}
Gunawardena, J., 2012. A linear framework for time-scale separation in
  nonlinear biochemical systems. PLoS ONE 7, e36321.

\bibitem[{HJ72}]{hj72}
Horn, F., Jackson, R., 1972. General mass action kinetics. Arch. Rational Mech.
  Anal. 47, 81--116.

\bibitem[{KVBAWF04}]{kvbawf}
Kramer, B.~P., Viretta, A.~U., Baba, M. D.-E., Aubel, D., Weber, W.,
  Fussenegger, M., 2004. An engineered epigenetic transgene switch in mammalian
  cells. Science 22, 867--70.

\bibitem[{LK99}]{lk99}
Laurent, M., Kellershohn, N., 1999. Multistability: a major means of
  differentiation and evolution in biological systems. Trends Biochem. Sci. 24,
  418--22.

\bibitem[{MG08}]{rg08}
Manrai, A., Gunawardena, J., 2008. The geometry of multisite phosphorylation.
  Biophys. J. 95, 5533--43.

\bibitem[{PMDSC12}]{psc11}
{P{\'e}rez Mill{\'a}n}, M., Dickenstein, A., Shiu, A., Conradi, C., 2012.
  Chemical reaction systems with toric steady states. Bull. Math. Biol. 74,
  1027--65.

\bibitem[{RS93}]{rs93}
Russo, F.~D., Silhavy, T.~J., 1993. The essential tension: opposed reactions in
  bacterial two-component regulatory systems. Trends Microbiol. 1, 306--10.

\bibitem[{SF10}]{sf09}
Shinar, G., Feinberg, M., 2010. Structural sources of robustness in biochemical
  networks. Science 327, 1389--91.

\bibitem[{SMMA07}]{smma07}
Shinar, G., Milo, R., Mart{\'i}nez, M.~R., Alon, U., 2007. Input-output
  robustness in simple bacterial signaling systems. Proc. Natl. Acad. Sci. USA
  104, 19931--5.

\bibitem[{SRA09}]{sra09}
Shinar, G., Rabinowitz, J.~D., Alon, U., 2009. Robustness in glyoxylate bypass
  regulation. PLoS Comp. Biol. 5, e1000297.

\bibitem[{TG09a}]{mg09}
Thomson, M., Gunawardena, J., 2009{{a}}. The rational parameterisation
  theorem for multisite post-translational modification systems. J. Theor.
  Biol. 261, 626--36.

\bibitem[{TG09b}]{mg07}
Thomson, M., Gunawardena, J., 2009{{b}}. Unlimited multistability in
  multisite phosphorylation systems. Nature 460, 274--7.

\bibitem[{XG11}]{xg11b}
Xu, Y., Gunawardena, J., 2011. Realistic enzymology for post-translational
  modification: zero-order ultrasensitivity revisited, submitted.

\end{thebibliography}


\end{document}